\documentclass[twocolumn,numberedappendix,twocolappendix]{openjournal}

\usepackage{xcolor}
\usepackage{textgreek}
\usepackage[utf8]{inputenc}
\usepackage[english]{babel}
\usepackage{amsmath}
\usepackage{bm}
\usepackage{physics}
\usepackage{afterpage}

\usepackage{hyperref}
\hypersetup{
    unicode,
    colorlinks=true,
    linkcolor=linkcolor,
    citecolor=linkcolor,
    filecolor=linkcolor,
    urlcolor=linkcolor,
}
\usepackage{cleveref}
\crefname{table}{Table}{Tables}
\crefname{figure}{Fig.}{Figs.}
\crefname{equation}{Eq.}{Eqs.}
\usepackage{color,colortbl}
\definecolor{linkcolor}{rgb}{0.0,0.3,0.5}
\usepackage{tensind}
\tensordelimiter{?}
\DeclareGraphicsExtensions{.bmp,.png,.jpg,.pdf}
\usepackage{verbatim}
\usepackage[normalem]{ulem}
\usepackage{orcidlink}
\usepackage{soul}
\usepackage{acro}

\DeclareAcronym{BORG}{short = \texttt{BORG}, long  = \textit{Bayesian Origin Reconstruction from Galaxies}}
\DeclareAcronym{CMB}{short = CMB, long  = cosmic microwave background}
\DeclareAcronym{tSZ}{short = tSZ, long  = thermal Sunyaev–Zel'dovich}
\DeclareAcronym{kSZ}{short = kSZ, long  = kinetic Sunyaev–Zel'dovich}
\DeclareAcronym{HMC}{short = HMC, long = Hamiltonian Monte Carlo}
\DeclareAcronym{NUTS}{short = NUTS, long = No-U-Turn Sampler}
\DeclareAcronym{LCDM}{short = $\Lambda$CDM, long  = $\Lambda$-cold dark matter}
\DeclareAcronym{AGN}{short = AGN, long = active galactic nucleus}
\DeclareAcronym{LUM}{short = LUM, long = Local Universe Model}
\DeclareAcronym{CIB}{short = CIB, long = cosmic infrared background}

\newcommand{\Mpch}{\ensuremath{h^{-1}\,\mathrm{Mpc}}}
\newcommand{\Mpc}{\ensuremath{\mathrm{Mpc}}}

\newcommand{\Msunh}{\ensuremath{h^{-1}\,M_\odot}}
\newcommand{\Msun}{\ensuremath{M_\odot}}
\newcommand{\kmsec}{\ensuremath{\mathrm{km}\,\mathrm{s}^{-1}}}
\newcommand{\kmsecMpc}{\ensuremath{\mathrm{km}\,\mathrm{s}^{-1}\,\mathrm{Mpc}^{-1}}}
\newcommand{\dex}{\ensuremath{\mathrm{dex}}}

\newcommand{\CBa}{\texttt{CSiBORG1}}
\newcommand{\CBb}{\texttt{CB2}}
\newcommand{\CBM}{\texttt{CBM}}
\newcommand{\TWOMPP}{2M\texttt{++}}

\graphicspath{{./figs/}}

\begin{document}

\title{Validating Digital Twins of the Local Universe with the Thermal Sunyaev-Zel'dovich Signal}

\author{Richard Stiskalek\orcidlink{0000-0002-0986-314X}}
\email{richard.stiskalek@physics.ox.ac.uk}
\affiliation{Astrophysics, University of Oxford, Denys Wilkinson Building, Keble Road, Oxford, OX1 3RH, UK}

\author{Harry Desmond\orcidlink{0000-0003-0685-9791}}
\affiliation{Institute of Cosmology \& Gravitation, University of Portsmouth, Dennis Sciama Building, Portsmouth, PO1 3FX, UK}

\begin{abstract}
The thermal Sunyaev–Zel'dovich (tSZ) effect provides a powerful probe of the thermal pressure of ionised gas in galaxy clusters and the cosmic web; constrained simulations reconstruct the mass and velocity fields of the local Universe.
We explore how these two may be mutually informative: the tSZ signal provides a benchmark for assessing the fidelity of constrained simulations, and constrained simulations contribute information on the positions, total masses and density profiles of cosmic web structures for use in tSZ studies.
We focus on cluster predictions in the \textit{Bayesian Origin Reconstruction from Galaxies} (\texttt{BORG}) paradigm, introducing \texttt{CSiBORG-Manticore}, a new state-of-the-art suite of digital twins---data-constrained posterior simulations whose initial conditions are inferred via Bayesian forward modelling.
We develop a framework for scoring constrained simulations on their ability to match measured \textit{Planck} Compton-$y$ maps around clusters, and use it to demonstrate improvement from previous \texttt{BORG} reconstructions.
We further validate halo masses against weak-lensing-calibrated X-ray masses from \textit{eROSITA}.
We also show how high-fidelity digital twins offer a practical route to extracting additional information from tSZ data through a novel calibration of the mass--observable relation, and provide a complementary framework to purely statistical analyses of Compton-$y$ maps.
This paves the way for integrating the large-scale structure information inherent in constrained simulations into the study of CMB secondary anisotropies.
\end{abstract}


\maketitle

\section{Introduction}\label{sec:intro}

The \ac{tSZ} effect is a premier tool in modern cosmology and cluster astrophysics.
It arises from inverse Compton scattering of \ac{CMB} photons by hot, ionised electrons, primarily in the intracluster medium of galaxy groups and clusters.
This scattering conserves photon number but redistributes photon energies, producing a characteristic spectral distortion of the \ac{CMB} that is independent of redshift.
As a result, the \ac{tSZ} effect provides a nearly redshift-independent probe of the integrated thermal energy of ionised gas along the line of sight, making it a powerful tracer of the cluster population, large-scale structure and baryonic physics~\citep{SunyaevZeldovich1970,SunyaevZeldovich1972,Mroczkowski2019}.

In the context of galaxy clusters, the integrated \ac{tSZ} signal provides a mass proxy because it traces the total thermal energy of ionised gas bound in the cluster potential well.
Observational analyses calibrate this $Y$–mass relation either empirically by combining \ac{tSZ} measurements with independent mass estimates (e.g.\ weak gravitational lensing or X-ray hydrostatic masses) or by fitting external scaling relations based on simulations and multi-wavelength data; the latter includes an implicit ``mass bias'' parameter quantifying the ratio of the observable-inferred mass to the true mass~\citep{vonderLinden2014,Hoekstra2015,Planck2016tSZ,Salvati2019}.
Mass estimates derived from the \ac{tSZ} signal are then propagated into cluster cosmology analyses of number counts and clustering, and compared to masses obtained from other techniques, enabling assessments of systematic biases and improvements in cosmological parameter constraints~\citep{Planck_2014_SZcounts,Planck_2016_SZcosmo,deHaan2016,Bocquet2019,Hilton2021,Bocquet2024a,Bocquet2024b}.

Compton-$y$ maps are also used beyond individual cluster masses~\citep{Planck_2016_tSZ,Chandran_2023,McCarthyHill2024}: they constitute a projected tracer of the large-scale electron pressure field, which can be statistically analysed through its auto-power spectrum or cross-correlations with other large-scale structure tracers such as galaxy density or weak gravitational lensing.
These cross-correlations measure the joint spatial distribution of baryons and matter, providing independent constraints on astrophysical processes and cosmology~\citep{Koukoufilippas2020,Yan2021,Troster2022,Koukoufilippas2022,LaPosta2025}.
Using such two-point statistics, one can test models of gas thermodynamics and feedback~\citep{Battaglia_1}, and relate the \ac{tSZ} signal to other mass tracers on cosmic scales, making Compton-$y$ maps a versatile probe in multi-probe analyses of structure formation.

The aim of the paper is to connect the \ac{tSZ} effect with constrained simulations of the local Universe and develop a framework within which each may be used to inform the other.
Constrained simulations offer a complementary approach to standard cosmological simulations by tailoring the simulation volume to the actual local Universe.
Rather than sampling a random realisation of \ac{LCDM}, these methods aim to reproduce the observed large-scale structure by incorporating data such as galaxy redshift surveys or peculiar velocity measurements into the reconstruction of initial conditions (the phases of the initial density modes).
This strategy has enabled detailed studies of the nearby cosmic web, the dynamics of the Local Group, and the properties of prominent clusters and superclusters, while retaining statistical consistency of the underlying cosmological model on unconstrained scales~\citep[e.g.][]{Dolag2005_CORUSCANT,Gottloeber2010_CLUES,Sorce2016_CLUES}.
A state-of-the-art approach to constructing the constraints is Bayesian forward-modelling as encapsulated in the \ac{BORG} paradigm, which provides a full posterior distribution of initial conditions consistent with the local galaxy number density field~\citep{Jasche_2013}.

The \ac{tSZ} effect is a particularly natural observable to combine with constrained simulations, as it directly probes the integrated thermal pressure of ionised gas within the same large-scale structures that constrained reconstructions seek to reproduce.
In contrast to purely statistical analyses of Compton-$y$ maps---such as power spectra, stacking measurements, or cluster number counts---constrained simulations retain the spatial phase information of the density field, enabling object-by-object comparison between observed and simulated clusters and their environments.
This allows explicit modelling of projection effects along the line of sight, identification of dominant contributors to the \ac{tSZ} signal in specific directions, and physically motivated interpretation of residuals that would otherwise be absorbed into effective scatter or nuisance parameters.
The simulations also provide a means of calibrating the mass--Compton-$y$ relation that does not require any other mass proxy.
Such considerations are particularly relevant in the local Universe, where a small number of massive clusters and supercluster environments dominate the \ac{tSZ} signal.

We use digital twins, constrained simulations that sample the Bayesian posterior of the local density field, to investigate how \ac{tSZ} observations can be interpreted and augmented by knowledge of the underlying large-scale structure.
Building on \ac{BORG}-based reconstructions, we construct digital twins of the nearby Universe that reproduce the observed positions of galaxy clusters with high fidelity, and compare their predicted thermal pressure distribution to observed Compton-$y$ maps.
Our analysis focuses on cluster matching, positional accuracy, and the correspondence between simulated and observed \ac{tSZ} signals, rather than on deriving new cosmological constraints.
By doing so, we demonstrate how digital twins provide a practical framework for extracting additional physical information from \ac{tSZ} data and for validating assumptions underlying standard \ac{tSZ} analyses, while remaining complementary to established statistical approaches.

The paper is structured as follows.
Section~\ref{sec:data} describes the simulation suites and Compton-$y$ maps employed, as well as the X-ray and optical cluster data used for further validation.
Section~\ref{sec:method} describes our methodology, including our approach for associating haloes between realisations in each suite and with Compton-$y$ hotspots, the metric we develop for scoring the simulations on the basis of the Compton-$y$ signal, and our method for inferring mass--mass and mass--observable scaling relations.
Section~\ref{sec:results} details the results, while Section~\ref{sec:discussion} provides a literature comparison, situates our validation metric within a broader context and proposes future directions.
Section~\ref{sec:conclusion} concludes.

\section{Data}\label{sec:data}

We describe the constrained simulation suites used in this work (Section~\ref{sec:digital_twins}), the \textit{Planck} \ac{tSZ} map and cluster catalogue (Sections~\ref{sec:planck_tsz_map} and~\ref{sec:planck_tsz_catalogue}), the eROSITA X-ray cluster catalogue (Section~\ref{sec:erosita_catalogue}), and the selection of nearby clusters for detailed analysis (Section~\ref{sec:named_clusters}).

\subsection{Digital twins}\label{sec:digital_twins}

We employ two suites of digital twins\footnote{While coming under the umbrella heading of ``constrained simulations'', these are more precisely described as data-constrained posterior simulations, reflecting that their initial conditions are inferred via a full Bayesian forward model with explicit uncertainty quantification.}---\texttt{CSiBORG2} (\CBb) and \texttt{CSiBORG-Manticore} (\CBM), the latter a new suite of digital twins introduced in this work---representing successive generations of reconstructions based on the \ac{BORG} algorithm applied to the \TWOMPP\ galaxy catalogue~\citep{Lavaux_2011}.
\ac{BORG} produces a posterior distribution of voxel densities at $z = 1000$ by forward-modelling the initial conditions with redshift-space distortions, selection effects and galaxy biasing, and comparing to the observed galaxy number density via a Poisson likelihood~\citep{Jasche_2013, Jasche_2015, Lavaux_2016, Leclercq_2017, Jasche_2019, Lavaux_2019, Porqueres_2019, Stopyra_2023, McAlpine2025_ManticoreCatalogue}.
Each suite comprises multiple $N$-body simulations, each representing a single draw from the \ac{BORG} posterior, thereby capturing uncertainties in the large-scale structure constraints while delivering high-resolution dark-matter simulations by adding random-phase small-scale modes consistent with the imposed large-scale density field and the \ac{LCDM} power spectrum.

In both suites, constraints are imposed within the \TWOMPP\ region, a spherical volume of radius ${\sim}155~\Mpch$ centred on the Milky Way ($z \lesssim 0.05$), embedded in a periodic box of side length $1000~\Mpc$.
Beyond this volume, the density field is largely unconstrained, serving only to provide the correct large-scale forces expected in \ac{LCDM}.
The \ac{BORG} density field is sampled on a $256^3$ grid with a resolution of $3.9~\Mpc$.
We now briefly describe each suite in turn.

\CBb\ is the second iteration of the \emph{Constrained Simulations in BORG} (\texttt{CSiBORG}) suite.
It adopts a similar setup to the original \texttt{CSiBORG}~\citep{Bartlett2021_CSiBORG} but uses the updated \ac{BORG} initial conditions of~\citet{Stopyra_2023}.
It comprises $20$ dark matter-only $N$-body runs within a $677.7~\Mpch$ box centred on the Milky Way.
The \ac{BORG} initial conditions employ a $20$-step \texttt{COLA} integrator~\citep{Tassev_2013_COLA} for improved accuracy, are linearly extrapolated to $z = 69$ and supplemented with small-scale white noise on a $2048^3$ mesh covering the constrained \TWOMPP\ volume.
This high-resolution zoom-in region, spanning the ${\sim}155~\Mpch$ radius where the \TWOMPP\ catalogue has highest completeness, yields a spatial resolution of $0.33~\Mpch$ and a particle mass of $3 \times 10^9~\Msunh$.
A $10~\Mpch$ buffer zone surrounds this high-resolution region to ensure a smooth transition to the coarser outer volume, preventing artificial discontinuities in the density field at the boundary.
The simulations are evolved to $z = 0$ using \texttt{Gadget4}~\citep{Gadget4}, adopting cosmological parameters from~\citet{Planck_2020} incorporating lensing and baryon acoustic oscillation data: $\Omega_{\rm m} = 0.3111$, $\sigma_8 = 0.8102$, $H_0 = 67.66~\kmsecMpc$, $n_{\rm s} = 0.9665$, $\Omega_{\rm b} = 0.049$.
We use $20$ posterior samples resimulated for~\citet{Stiskalek_2025_VFO}.

The \CBM\ suite uses the \texttt{Manticore-Local} initial conditions from~\citet{McAlpine_2025}, representing the most recent \ac{BORG} reconstruction, resimulated with the \texttt{CSiBORG} zoom-in setup.
We use $50$ posterior samples, drawn from four independent chains, resimulated with \texttt{Gadget4} using the same zoom-in parameters as \CBb, yielding a particle mass of $3 \times 10^9~\Msunh$.
The cosmological parameters are drawn from the Dark Energy Survey Year 3 ``$3 \times 2$pt + All Ext.'' \ac{LCDM} analysis~\citep{DES_Y3}: $h=0.681$, $\Omega_{\rm m} = 0.306$, $\Omega_\Lambda = 0.694$, $\Omega_{\rm b} = 0.0486$, $A_{\rm s} = 2.099 \times 10^{-9}$, $n_{\rm s} = 0.967$, and $\sigma_8 = 0.807$.
\citet{McAlpine_2025} introduced these initial conditions along with the \texttt{Manticore-Local} suite, in which they are resimulated at a uniform $1024^3$ resolution, and showed that it surpasses earlier reconstructions in velocity field tests, exhibits good agreement with observed cluster masses and positions, and produces a $z = 0$ power spectrum and halo mass function that are in excellent agreement with \ac{LCDM} expectations.
We adopt the \CBM\ suite rather than \texttt{Manticore-Local} for convenience; while it offers higher particle resolution, this makes no significant difference for the massive haloes studied here.

\subsection{\textit{Planck} tSZ map}\label{sec:planck_tsz_map}

We use the \textit{Planck} all-sky Compton-$y$ map as our primary observational dataset for validating the \ac{tSZ} signal in the constrained simulations.
The amplitude of the \ac{tSZ} effect is quantified by the Compton-$y$ parameter, defined as
\begin{equation}
    y \equiv \frac{\sigma_{\rm T}}{m_{\rm e} c^2} \int n_{\rm e}(l) \, k_{\rm B} T_{\rm e}(l) \, \mathrm{d}l,
\end{equation}
where $\sigma_{\rm T}$ is the Thomson cross-section, $m_{\rm e}$ the electron mass, $n_{\rm e}$ and $T_{\rm e}$ are the electron number density and temperature, respectively, and the integral is taken along the line of sight $l$.
The Compton-$y$ parameter is therefore proportional to the integrated electron pressure,
\begin{equation}
    P_{\rm e} = n_{\rm e} k_{\rm B} T_{\rm e},
\end{equation}
and directly traces the thermal energy content of ionised gas, with minimal sensitivity to gas clumping compared to X-ray observables.
Observationally, the \ac{tSZ} signal is extracted from multi-frequency \ac{CMB} data by exploiting its distinctive spectral dependence, which produces a decrement in \ac{CMB} intensity below ${\simeq}217\,\mathrm{GHz}$ and an increment above it.
Component-separation techniques applied to multi-band data from experiments such as \textit{Planck}, ACT, or SPT are used to isolate the \ac{tSZ} contribution and construct full-sky or partial-sky Compton-$y$ maps~\citep{Planck2016tSZ}.

We use the all-sky Compton-$y$ parameter map constructed from the \textit{Planck} PR4 data release by~\citet{McCarthyHill2024}.
This map was produced using a Needlet Internal Linear Combination (NILC) pipeline applied to the nine \textit{Planck} frequency channels from 30 to 857~GHz.
The map employs a moment-based deprojection method to minimise contamination from the \ac{CIB}, which otherwise biases the recovered Compton-$y$ signal at small angular scales.
The map is consistent with the earlier \textit{Planck} release~\citep{Planck_2016_tSZ} on large scales, but has 10--20\% lower noise on small scales.

\subsection{\textit{Planck} tSZ cluster catalogue}\label{sec:planck_tsz_catalogue}

We use the second \textit{Planck} catalogue of Sunyaev–Zel'dovich sources (PSZ2;~\citealt{Planck_2016_PSZ2})\footnote{\url{https://heasarc.gsfc.nasa.gov/w3browse/all/plancksz2.html}}, which contains 1653 detections, of which 1203 are confirmed clusters with identified counterparts; we restrict our analysis to the latter.
The catalogue contains the integrated Compton-$y$ parameter $Y_{5R_{500}}$ for each source, computed via aperture photometry within an angular radius corresponding to $5 R_{500{\rm c}}$.
The cluster size $R_{500{\rm c}}$ is derived iteratively using a $Y_{500{\rm c}}$--$\theta_{500{\rm c}}$ prior combined with the $Y_{500{\rm c}}$--$M_{500{\rm c}}$ scaling relation from~\citet{Planck_2014_SZcounts}, itself calibrated to X-ray hydrostatic masses from the MCXC~\citep{Piffaretti_2011}.

We convert $Y_{5R_{500}}$ to $Y_{500{\rm c}}^{\rm tSZ}$ using $Y_{5R_{500}} = 1.81 Y_{500{\rm c}}^{\rm tSZ}$, derived under the spherical assumption with the universal pressure profile~\citep{Arnaud_2010,Planck_2011_ESZ}.
We then convert to physical units by multiplying by $D_{\rm A}^2(z)$, where $D_{\rm A}(z)$ is the angular diameter distance computed from the observed heliocentric redshift.
Finally, we apply the self-similar scaling by multiplying by $E(z)^{-2/3}$, where
\begin{equation}
    E(z) = \sqrt{\Omega_{\rm m}(1+z)^3 + 1 - \Omega_{\rm m}},
\end{equation}
though this correction is negligible at $z < 0.05$.
We use these measurements alongside the \ac{tSZ} mass estimates reported in the catalogue to compare with predictions from the digital twin halo masses.
Note that the reported masses are likely biased low by ${\sim}30\%$ with respect to weak lensing masses~\citep{Sereno_2017}, and \textit{Planck} cluster counts using \ac{tSZ} masses yield a relatively low $\sigma_8$ compared to the fiducial \textit{Planck} \ac{CMB} analysis~\citep{Planck_2016_SZcosmo,Planck_2020_cosmo}.
We therefore prefer to compare directly to $Y_{500{\rm c}}^{\rm tSZ}$ rather than the derived masses, which may not be reliable, though we shall also show the mass comparison.
We select clusters with $M_{500{\rm c}}^{\rm tSZ} > 10^{14}~\Msun$ and $z < 0.05$, yielding 56 clusters for our analysis.
\vspace{1em}

\subsection{eROSITA X-ray cluster catalogue}\label{sec:erosita_catalogue}

We employ the eROSITA X-ray cluster catalogue primarily because its masses are calibrated via state-of-the-art weak gravitational lensing, providing an independent mass reference against which to compare the digital twin halo masses.
We use the first eROSITA All-Sky Survey cluster catalogue~\citep{Bulbul2024_eRASS1}, which contains 12,247 optically confirmed galaxy clusters and groups detected as extended X-ray sources in the western Galactic hemisphere.
The catalogue covers 13,116~deg$^2$ observed during eROSITA's first six months of operation aboard the Spectrum-Roentgen-Gamma satellite~\citep{Predehl2021_eROSITA}.
For each cluster, the catalogue reports X-ray properties including flux, luminosity $L_{500{\rm c}}^{\rm X-ray}$, and temperature, as well as derived quantities such as total mass $M_{500{\rm c}}^{\rm X-ray}$ and gas mass~\citep[see also][for X-ray scaling relations from eFEDS]{NguyenDang2024}.

Crucially, the eROSITA cluster masses are calibrated using weak gravitational lensing measurements from the Dark Energy Survey Year~3 data~\citep{Grandis2024_eROSITAWL}.
This calibration determines the scaling between X-ray count rate and halo mass by measuring the lensing signature in background galaxy shapes caused by eROSITA-selected clusters, using a Bayesian population model that accounts for sample selection, contamination and instrumental systematics.
Applying the same selection criteria ($M_{500{\rm c}}^{\rm X-ray} > 10^{14}~\Msun$ and $z < 0.05$) yields 74 clusters in this case.

\subsection{Selected nearby clusters}\label{sec:named_clusters}

We select 18 well-known nearby clusters for detailed analysis, listed in~\cref{tab:named_clusters}.
These clusters span a redshift range $0.005 \lesssim z \lesssim 0.050$ and are distributed across the sky, covering prominent superclusters including Hydra--Centaurus, Hercules, Coma, and Shapley.
Clusters are labelled by their host supercluster where applicable, with their Abell catalogue designation~\citep{Abell1958} in parentheses; multiple clusters sharing a supercluster name (e.g.\ Hercules) are distinct systems within the same optically identified large-scale overdensity, not necessarily gravitationally bound to one another.
For each cluster, we report the observed Galactic coordinates $(\ell, b)$ and \ac{CMB}-frame recession velocity $cz_{\rm CMB}$ obtained from NED\footnote{\url{http://ned.ipac.caltech.edu}}, along with masses from the \textit{Planck} \ac{tSZ} and eROSITA X-ray catalogues matched by angular separation and redshift.
Where both masses are available, the weak-lensing-calibrated eROSITA masses are systematically higher than the \ac{tSZ} masses, reflecting a known bias in the \textit{Planck} \ac{tSZ} mass calibration.

\begin{table*}
    \centering
    \renewcommand{\arraystretch}{1.15}
    \caption{Selected nearby clusters for detailed analysis, ordered by recession velocity. We report the Galactic longitude $\ell$, Galactic latitude $b$, \ac{CMB}-frame recession velocity $cz_{\rm CMB}$, the \textit{Planck} \ac{tSZ} mass~\citep{Planck_2016_PSZ2}, and the weak-lensing-calibrated eROSITA mass~\citep{Bulbul2024_eRASS1}, where available. Virgo lacks a \ac{tSZ} mass due to its large angular extent relative to the \textit{Planck} beam; other clusters may be absent due to proximity to the Zone of Avoidance. Some clusters lack eROSITA masses due to the half-sky coverage of the eRASS1 data release.}
    \label{tab:named_clusters}
    \begin{tabular}{lccccc}
        \hline
        Cluster & $cz_{\rm CMB}$ [\kmsec] & $\ell$ [deg] & $b$ [deg] & $\log M_{500{\rm c}}^{\rm tSZ}$ [$\Msun$] & $\log M_{500{\rm c}}^{\rm X-ray}$ [$\Msun$] \\
        \hline
        Virgo             &  1636 & 283.8 &  74.4 & ---   & --- \\
        Centaurus (A3526) &  3403 & 302.4 &  21.6 & 14.12 & 14.43 \\
        Hydra (A1060)     &  4058 & 269.6 &  26.5 & ---   & 14.20 \\
        Norma (A3627)     &  4955 & 325.3 & $-7.1$ & 14.38 & 14.61 \\
        Perseus (A426)    &  4995 & 150.6 & $-13.3$ & ---   & --- \\
        Leo (A1367)       &  6890 & 235.1 &  73.0 & 14.22 & 14.55 \\
        Coma (A1656)      &  7463 &  58.1 &  88.0 & 14.86 & --- \\
        Hercules (A2199)  &  9113 &  62.9 &  43.7 & 14.46 & --- \\
        Abell 496         &  9849 & 209.6 & $-36.5$ & 14.43 & 14.83 \\
        Hercules (A2063)  & 10634 &  12.9 &  49.7 & 14.28 & --- \\
        Hercules (A2151)  & 11024 &  31.6 &  44.5 & ---   & --- \\
        Hercules (A2147)  & 11072 &  29.0 &  44.5 & 14.55 & --- \\
        Shapley (A3571)   & 11965 & 316.3 &  28.6 & 14.67 & 14.82 \\
        Abell 548         & 12363 & 230.3 & $-24.8$ & ---   & 13.68 \\
        Abell 119         & 13004 & 125.7 & $-64.1$ & 14.54 & --- \\
        Abell 1736        & 13823 & 312.6 &  35.0 & 14.46 & --- \\
        Abell 1644        & 14448 & 304.9 &  45.5 & 14.57 & 14.59 \\
        Shapley (A3558)   & 14784 & 312.0 &  30.7 & 14.68 & 14.81 \\
        \hline
    \end{tabular}
\end{table*}

\section{Methodology}\label{sec:method}

We describe our methodology for comparing observed \ac{tSZ} signals with predictions from constrained simulations.
In Section~\ref{sec:tsz_signal}, we detail how we measure the \ac{tSZ} signal at halo positions, which is used for our named-cluster catalogue.
In Section~\ref{sec:stacked_profiles}, we describe stacked radial profiles, which are used to test the mass dependence of the signal.
In Section~\ref{sec:associations}, we introduce halo associations, which link the same halo across realisations.
In Sections~\ref{sec:classical_matching} and~\ref{sec:lum_matching}, we describe the matching procedures used to link catalogues of observed clusters to these associations.
Finally, in Section~\ref{sec:scaling_relations}, we describe the Bayesian linear regression used to infer scaling relations while marginalising over the mass uncertainty.

\subsection{\ac{tSZ} signal at halo positions}\label{sec:tsz_signal}

We measure the \ac{tSZ} signal of each halo using aperture photometry on a Compton-$y$ map in \texttt{HEALPix}\footnote{\url{https://healpix.sourceforge.io}} format~\citep{healpix}.
For a given pointing and aperture radius $\theta_{\rm ap}$, we use the \texttt{healpy}\footnote{\url{https://healpy.readthedocs.io/en/latest/}} implementation~\citep{healpy} to identify all \texttt{HEALPix} pixels within the circular aperture.
We apply a mask to exclude invalid regions, including the Galactic plane, point sources, and pixels marked as unseen.
The enclosed signal is then computed as the arithmetic mean of the Compton-$y$ values over all valid pixels within the aperture.
For each halo, we set the aperture radius equal to $2\theta_{500{\rm c}}$, where $\theta_{500{\rm c}}$ is the angular radius corresponding to $R_{500{\rm c}}$ at the halo's comoving distance and $R_{500{\rm c}}$ is the radius within which the mean density is 500 times the critical density.

To assess the significance of the measured signal, we compute an empirical $p_{\rm tSZ}$-value by comparing it to a null distribution obtained from random sky positions.
We measure the enclosed signal at $N_{\rm rand} = 10^5$ random pointings, excluding the Galactic plane, at the same aperture size as the halo.
The halo's signal $S$ is then ranked within this distribution, and the $p_{\rm tSZ}$-value is
\begin{equation}
    p_{\rm tSZ} = 1 - \frac{{\rm rank}(S)}{N_{\rm rand}}.
\end{equation}
This $p_{\rm tSZ}$ represents the fraction of random pointings with a signal equal to or exceeding $S$, i.e.~the probability of observing a signal as high as $S$ by chance.
A low $p_{\rm tSZ}$ therefore indicates that the measured signal is significantly higher than expected from random sky positions.
$p_{\rm tSZ}$ is sensitive to the halo's angular position on the sky, as different regions exhibit varying noise and foreground contamination levels, and partly to its physical size via the aperture radius.

\subsection{Stacked 1D radial profiles}\label{sec:stacked_profiles}

To construct average radial profiles across multiple sources, we stack individual profiles measured at normalised radii.
For each source with angular size $\theta_{500{\rm c}}$, we measure the enclosed signal at normalised angular separations $\theta / (2\theta_{500{\rm c}})$, yielding a radial profile that can be compared across sources of different angular sizes arising from their varying distances and masses.
The stacked profile is then computed as the mean across all sources at each normalised radius, with uncertainties estimated via bootstrap resampling.
To assess the significance of the stacked signal, we construct random stacks by measuring profiles at random sky positions, with aperture sizes resampled from the distribution of source angular sizes.
Each random stack contains the same number of profiles as the data stack, and we repeat this process many times to build a null distribution.
\vspace{2em}

\subsection{Halo associations}\label{sec:associations}

Halo associations were introduced by~\citet{McAlpine2025_ManticoreCatalogue} to identify the ``same'' halo across realisations of the \ac{BORG} posterior represented by the suite of digital twins.
An association is a set of haloes (at most one per realisation) that reside at approximately the same position at $z=0$ and have similar masses.
Following~\citet{McAlpine2025_ManticoreCatalogue}, we identify associations using the \texttt{DBSCAN} clustering algorithm with linking length $\epsilon = 1.75~\Mpch$ and a minimum membership of nine.
When multiple haloes from the same realisation fall into the same cluster, we retain only the one closest to the centroid (the mean position of all cluster members), since associations are defined to contain at most one halo per realisation.
We additionally discard any outliers whose masses deviate by more than $0.3~\dex$ from the mean logarithmic mass of the association members, and recalculate the centroid from the surviving members.
Each association is furthermore characterised by $f_{\rm present}$, defined as the number of member haloes divided by the total number of realisations.

\subsection{Classical matching}\label{sec:classical_matching}

We match observed clusters from the \textit{Planck} \ac{tSZ} catalogue to halo associations using a classical angular separation and redshift criterion.
For each observed cluster-association pair, we compute the angular separation on the sky and the line-of-sight velocity difference $\Delta cz = |cz_{\rm obs} - cz_{\rm sim}|$ in the \ac{CMB} frame, where $cz_{\rm obs}$ and $cz_{\rm sim}$ are the recessional velocities of the observed cluster and simulated halo, respectively.
The redshift of the simulated halo is computed as
\begin{equation}
    1 + z_{\rm sim} = (1 + z_{\rm cosmo})(1 + z_{\rm pec}),
\end{equation}
where $z_{\rm cosmo}$ is the cosmological redshift corresponding to the halo's comoving distance and $z_{\rm pec} = V_{\rm los} / c$ is the Doppler shift from its peculiar velocity $V_{\rm los}$ along the line of sight.
A pair is considered a valid candidate if $\Delta\theta < 60^\prime$ and $\Delta cz < 300~\kmsec$.
We further require that at least 50\% of the association's member haloes satisfy both criteria, ensuring robust matches even for associations with significant positional scatter.
We opt for relatively stringent matching criteria to minimise spurious matches, which would tend to flatten the inferred scaling relation slopes; we note that additional halo--cluster pairs lie just beyond these thresholds and would likely constitute valid matches.

Among the valid candidate pairs, we assign matches using a greedy algorithm that minimises the three-dimensional redshift-space comoving distance between the observed cluster position and the association centroid.
At each iteration, the algorithm selects the pair with the smallest distance, assigns the match, and removes both the cluster and association from further consideration.
This process continues until no valid pairs remain.

\subsection{LUM matching}\label{sec:lum_matching}

For our set of 18 named nearby clusters (described in Section~\ref{sec:data}), we employ the \ac{LUM} matching procedure of~\citet{LUM}, which quantifies the significance of each cluster-halo association.
We note that the classical matching procedure would yield very similar pairs; however, the LUM approach provides additional insight into the match significance.
For a halo of mass $M_{200{\rm c}}$ at distance $r$ from the observed cluster position, we compute the cumulative probability of finding such a halo within that distance under the assumption of a uniform spatial distribution.
Here $M_{200{\rm c}}$ is the mass enclosed within $R_{200{\rm c}}$, the radius within which the mean density is 200 times the critical density of the Universe.
The probability density of finding the nearest halo at distance $r$ is
\begin{equation}
    p(r) = 4 \pi r^2 n(M_{200{\rm c}}) \exp\left[ -\frac{4}{3} \pi r^3 n(M_{200{\rm c}}) \right],
\end{equation}
where $n(M_{200{\rm c}})$ is the halo mass function evaluated at $z = 0$ using the \texttt{colossus} package~\citep{colossus} with the model of~\citet{Tinker_2008}.
The cumulative distribution function, representing the probability of finding a halo within distance $r$, is
\begin{equation}
    p_{\rm LUM} = 1 - \exp\left[ -\frac{4}{3} \pi r^3 n(M_{200{\rm c}}) \right].
\end{equation}
A low value of $p_{\rm LUM}$ indicates that the halo is closer than expected by chance, suggesting a significant match; $p_{\rm LUM}$ can thus be interpreted as a $p$-value under the null hypothesis that the halo--cluster proximity arose by chance.
Note that $p_{\rm LUM}$ depends on both the halo mass and distance: more massive haloes are rarer and thus a match at a given distance is more significant.
For each observed cluster, we compute $p_{\rm LUM}$ for all haloes within a maximum search radius and select the match with the lowest value, subject to a threshold on the maximum allowed $p_{\rm LUM}$.
Since each association contains multiple haloes from different realisations, we compute $p_{\rm LUM}$ for each member halo, yielding a distribution of $p_{\rm LUM}$-values that characterises the match quality across the ensemble.
To summarise the match quality for a cluster-association pair, we take the median $p_{\rm LUM}$ across all member haloes.

To assign matches, we first compute the median $p_{\rm LUM}$ over member haloes for each observed cluster-association pair, forming a matrix.
We then apply a greedy algorithm: at each iteration, we select the pair with the lowest median $p_{\rm LUM}$, assign the match, and remove both the cluster and association from further consideration.
The process continues until no pairs remain below a specified threshold.

\subsection{Scaling relation inference}\label{sec:scaling_relations}

To infer scaling relations between catalogue quantities (e.g., $\log Y_{500{\rm c}}^{\rm tSZ}$ or calibrated mass) and \ac{BORG} halo masses, we use a linear regression model that accounts for the uncertainty in the simulated masses, which are the independent variable.
For each matched association, we have a distribution of halo masses from the member haloes across realisations, rather than an estimate with Gaussian uncertainty.
We infer a linear model
\begin{equation}
    y = m \log \left( \frac{M_{500{\rm c}}^{\texttt{BORG}}}{10^{14} \Msunh} \right) + c,
\end{equation}
where $y$ is the logarithm of the catalogue quantity with Gaussian error $\sigma_y$, $M_{500{\rm c}}^{\texttt{BORG}}$ is the halo mass from \ac{BORG} (the mass enclosed within $R_{500{\rm c}}$), $m$ is the slope, and $c$ is the intercept.
We similarly normalise the dependent variable: for catalogue masses we use $y = \log (M_{500{\rm c}} / (10^{14}~\Msunh))$, for X-ray luminosities $y = \log (L_{500} / (10^{44}~{\rm erg\,s^{-1}}))$, and for the integrated Compton-$y$ parameter $y = \log (Y_{500{\rm c}}^{\rm tSZ} / (10^{-4}~{\rm Mpc}^2))$.
Pivoting the variables near their mean values reduces covariance between $m$ and $c$.
We also include an intrinsic scatter $\sigma_{\rm int}$ as a free parameter, which is added in quadrature to the observational error.

To account for the mass uncertainty, we marginalise over the mass samples.
Assuming that the matched clusters are independent, the total likelihood is the product of individual likelihoods.
For each observed cluster $i$ with mass samples $\{M_i^{(k)}\}_{k=1}^{K}$ from its matched association, the likelihood is
\begin{equation}
    \mathcal{L}_i = \frac{1}{K} \sum_{k=1}^{K} \mathcal{N}\left(y_i \mid m \log \frac{M_i^{(k)}}{10^{14} \Msunh} + c,\, \sqrt{\sigma_{y,i}^2 + \sigma_{\rm int}^2}\right),
\end{equation}
where $\mathcal{N}(y \mid \mu, \sigma)$ is a probability density function of a Gaussian distribution with mean $\mu$ and standard deviation $\sigma$.
We adopt flat priors on $m$, $c$, and $\sigma_{\rm int}$.
To sample the posterior distribution, we use the \acl{NUTS} (\ac{NUTS};~\citealt{Hoffman_2011}) method of \ac{HMC} as implemented in the \texttt{numpyro}\footnote{\url{https://num.pyro.ai/en/latest/}} package~\citep{Phan_2019, Bingham_2019}, collecting approximately $10^4$ effective samples and ensuring a Gelman--Rubin statistic $\hat{R}-1 \leq 0.001$ for convergence~\citep{Gelman_1992}.

To quantify the monotonic correlation between predicted and observed quantities, we compute the Spearman rank correlation coefficient $\rho_{\rm s}$.
We estimate its uncertainty by resampling: in each of $10^4$ iterations, we draw $x$ and $y$ from Gaussian distributions centred on their measured values with standard deviations equal to their uncertainties, and compute $\rho_{\rm s}$ for the resampled data.
We report the median and standard deviation of the resulting distribution.

We then calculate a significance $\tau$ at which the inferred scaling relation is in tension with theoretical expectations.
Under the self-similar model of cluster formation and hydrostatic equilibrium, the integrated Compton-$y$ signal within a characteristic radius (often $R_{500}$) scales with halo mass as $Y \propto M^{m} E^{2/3}(z)$, where $E(z)\equiv H(z)/H_0$ and $m\sim5/3$.
For soft-band X-ray luminosity, the self-similar expectation is $L_{\rm X} \propto M^{4/3}$, applicable to eROSITA's $0.5$--$2~{\rm keV}$ band.
This assumes purely gravitational cluster formation, hydrostatic and virial equilibrium, constant gas fraction and the non-relativistic limit~\citep{Kaiser_1986,Arnaud_2010}.
We therefore compute the marginal tension for the observable--mass relations as $\tau_m = |\bar{m} - m_{\rm ss}| / \sigma_m$, where $\bar{m}$ is the posterior mean, $\sigma_m$ is the posterior standard deviation of the slope, and $m_{\rm ss}$ is the self-similar expectation ($5/3$ for \ac{tSZ}, $4/3$ for X-ray).
For mass--mass relations, where the expectation is a one-to-one correspondence ($m = 1$, $c = 0$)---the latter following from pivoting both variables at the same value---we compute a joint tension using the Mahalanobis distance: $d^2 = (\bm{\theta} - \bm{\mu})^{\rm T} \bm{\Sigma}^{-1} (\bm{\theta} - \bm{\mu})$, where $\bm{\theta} = (1, 0)^{\rm T}$ is the target, $\bm{\mu}$ is the posterior mean of $(m, c)$, and $\bm{\Sigma}$ is the posterior covariance matrix.
We convert $d^2$ to a $p$-value via the $\chi^2$ distribution with two degrees of freedom, $p = 1 - F_{\chi^2}(d^2; 2)$, and then to an equivalent Gaussian significance $\tau_{m,c} = \Phi^{-1}(1 - p/2)$, where $\Phi^{-1}$ is the inverse standard normal cumulative distribution function.

\section{Results}\label{sec:results}

We present our validation of constrained simulations against \ac{tSZ} observations in three parts: per-cluster angular positioning tests (Section~\ref{sec:scoring_angular}), stacked radial profiles as a function of halo mass (Section~\ref{sec:scoring_stacked}), and mass scaling relations for matched cluster samples (Section~\ref{sec:scoring_mass_agreement}).

\subsection{Scoring angular position}\label{sec:scoring_angular}

\begin{figure*}
    \centering
    \includegraphics[width=\textwidth]{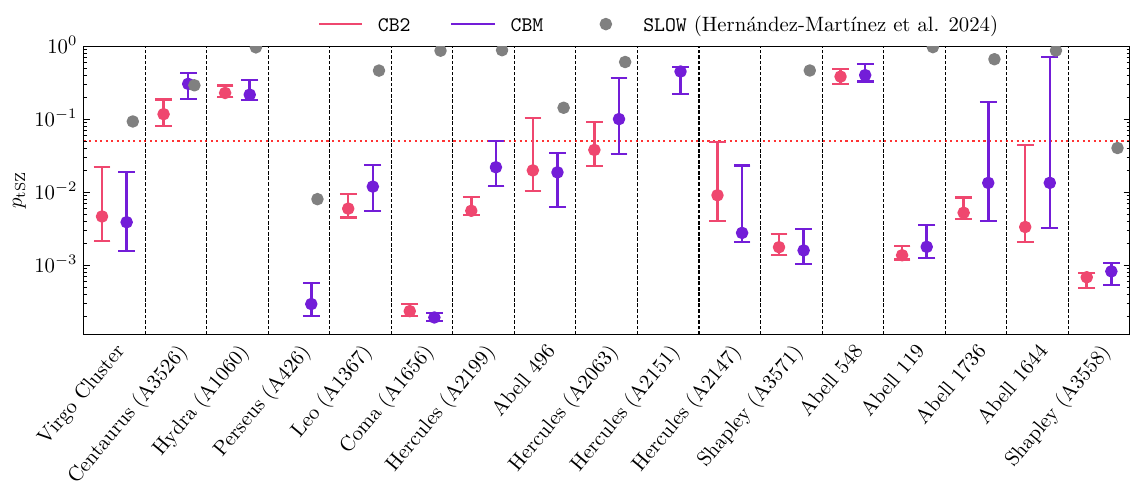}
    \caption{Distribution of $p_{\rm tSZ}$ for selected clusters from~\cref{tab:named_clusters} across realisations in \CBb\ and \CBM. For each cluster, $p_{\rm tSZ}$ is the fraction of random sky positions that yield an equal or higher \ac{tSZ} signal than the observed cluster position; low values indicate that the simulated halo lies at the observed \ac{tSZ} hotspot, indicative of a well-reconstructed cluster. The horizontal dashed line marks $p_{\rm tSZ} = 0.05$. We also show $p_{\rm tSZ}$ values for \texttt{SLOW} clusters~\citep{HernandezMartinez2024_SLOW}, computed using the same procedure with their simulated halo positions.}
    \label{fig:cluster_ranking}
\end{figure*}

We assess the angular positioning of simulated haloes by computing the \ac{tSZ} signal at each observed cluster position and comparing it against random sky locations to derive a $p$-value, $p_{\rm tSZ}$, for each cluster in each realisation.
For the majority of clusters, we find $p_{\rm tSZ} < 0.05$ across all realisations, indicating that the simulations place haloes at the correct angular positions (\cref{fig:cluster_ranking}).
The results are broadly consistent between \CBb\ and \CBM, with both simulations yielding significant detections for most selected clusters.
Coma and Shapley (A3558) are among the best-reproduced clusters in both simulations, with $\tilde{p}_{\rm tSZ} < 10^{-3}$.

\begin{figure*}
    \centering
    \includegraphics[width=0.95\textwidth]{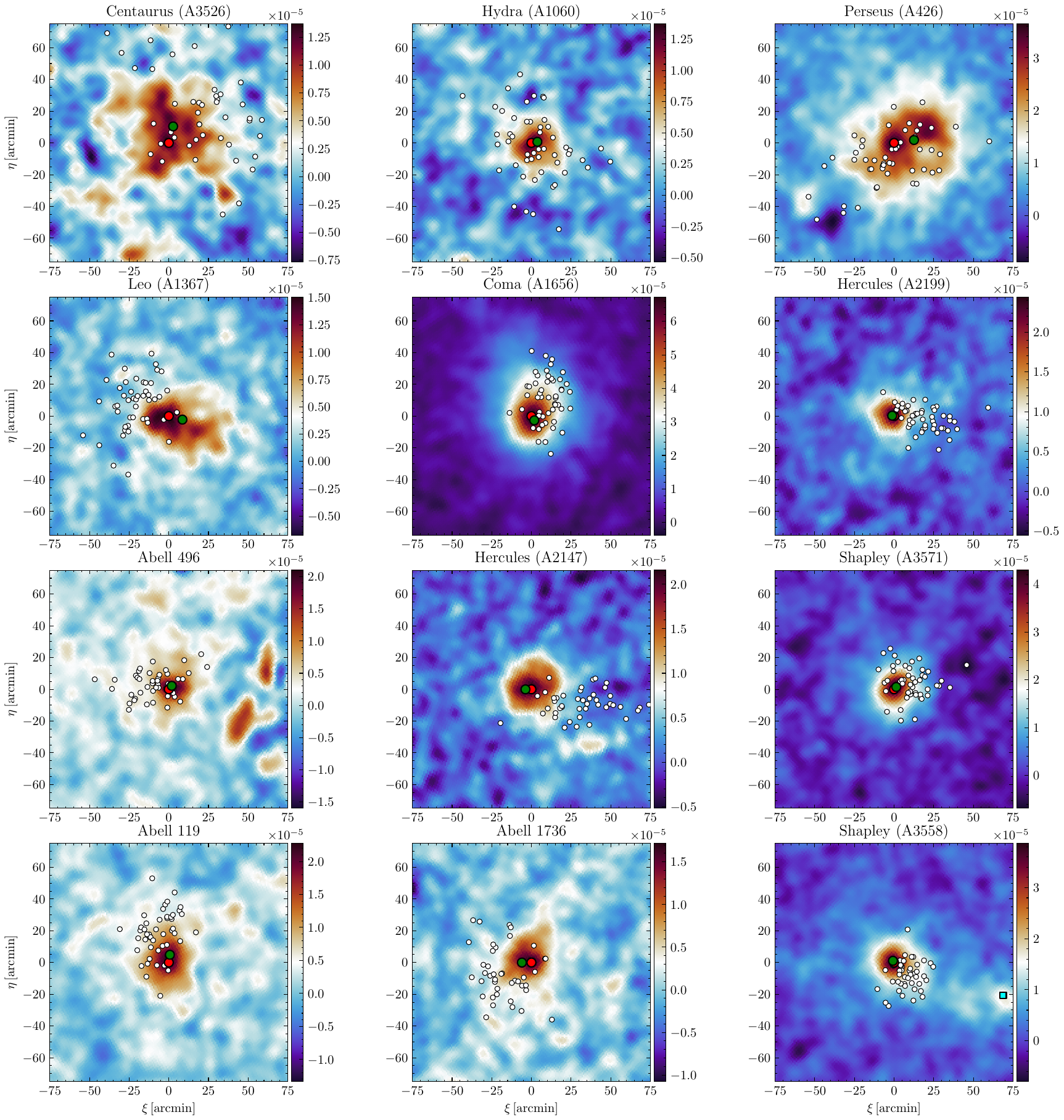}
    \caption{Compton-$y$ map cutouts centred on selected clusters. White dots mark the positions of \CBM\ halo realisations, the green dot indicates the reported cluster centre from the optical catalogue, and the red dot marks the local peak of the \ac{tSZ} map. For Shapley (A3558), the cyan dot marks the position of Shapley (A3562), which lies within the field of view.}
    \label{fig:Manti_clusters}
\end{figure*}

A notable exception is Perseus, for which we find no counterpart in \CBb\ but which is among the best-reproduced clusters in \CBM\ ($\tilde{p}_{\rm tSZ} = 2.9 \times 10^{-4}$).
Perseus is the archetypal cool-core cluster, with a sharply peaked X-ray surface brightness, short central cooling time, and \acl{AGN} feedback from NGC~1275~\citep{Fabian2006_Perseus}.
The compact, centrally concentrated \ac{tSZ} signal characteristic of cool cores is evident in~\cref{fig:Manti_clusters}.
The presence of Perseus in \CBM\ but not \CBb\ reflects the improved \ac{BORG} inference underlying the \texttt{Manticore} initial conditions~\citep{McAlpine_2025}, which employs an updated prior on the initial white noise field and new galaxy bias modelling.
The stability of the Perseus halo across the \CBM\ ensemble is quantified by $f_{\rm present} = 1.00$ (\cref{tab:cluster_properties}), indicating that it is robustly reproduced in all 50 posterior realisations.
Conversely, Hercules (A2151) is poorly reproduced in \CBM\ ($\tilde{p}_{\rm tSZ} = 0.45$, $f_{\rm present} = 0.54$) while no counterpart is found in \CBb.
Norma is well reproduced in both simulations ($p_{\rm LUM} = 2.1 \times 10^{-2}$ in \CBb\ and $8.7 \times 10^{-5}$ in \CBM); however, its low Galactic latitude ($b \approx -7^\circ$) places it within the masked region of the Compton-$y$ map, precluding a $p_{\rm tSZ}$ assessment.
We note that \CBM\ yields systematically higher halo masses than \CBb\ for the matched clusters (\cref{tab:cluster_properties}); we defer discussion of these mass differences to Section~\ref{sec:scoring_mass_agreement}.

\Cref{fig:Manti_clusters} shows \ac{tSZ} map cutouts for selected clusters, illustrating the angular correspondence between the \CBM\ haloes and the observed signal.
We exclude Norma (masked), Virgo (nearby with extended \ac{tSZ} signature), and Shapley (A3562), Hercules (A2151 and A2063), Abell~1644, and Abell~548 (all with relatively high $\tilde{p}_{\rm tSZ}$ values) from this visualisation (\cref{tab:cluster_properties}).

In both Shapley and Hercules, the \CBM\ halo positions exhibit systematic offsets towards neighbouring clusters, possibly indicating that \ac{BORG} merges clusters along filamentary structures into a single halo.
Within Shapley, halo positions are offset from the \ac{tSZ} and optical centres of A3558 towards A3562.
Dynamical mass estimates from~\citet{Haines2018_ShapleyFilaments} yield $M_{\rm dyn} = (14.8 \pm 1.4) \times 10^{14}~\Msun$ for A3558 and $(6.6 \pm 0.8) \times 10^{14}~\Msun$ for A3562, while weak lensing measurements by~\citet{Higuchi_2020} give $M_{200{\rm c}} = 4.5^{+2.8}_{-2.4} \times 10^{14}~\Msunh$ and $2.0^{+2.7}_{-1.8} \times 10^{14}~\Msunh$, respectively.
A3558 is a dynamically disturbed post-merger system, with X-ray observations revealing cold fronts extending to ${\sim}1.2~\rm Mpc$~\citep{Rossetti2007_A3558, Mirakhor2023_A3558}.
The complex may itself be the remnant of a cluster-cluster collision within the Shapley core~\citep{Bardelli1998_A3558}, implying that hydrostatic mass estimates are potentially biased.
Within Hercules, halo positions are similarly oriented towards A2151; this is consistent with A2147 being the dominant mass concentration, as dynamical estimates from~\citet{MonteiroOliveira_2022} find A2147 ($M \approx 13.5 \times 10^{14}~\Msun$) to be nearly five times more massive than A2151 ($M \approx 2.9 \times 10^{14}~\Msun$).

For comparison, we apply the same $p_{\rm tSZ}$ procedure to clusters from the \texttt{SLOW} constrained hydrodynamical simulation~\citep{HernandezMartinez2024_SLOW}, using the simulated halo positions reported in their Table~A.1.
Of the clusters in our sample (\cref{tab:named_clusters}), only Perseus ($p_{\rm tSZ} = 0.008$) and Shapley (A3558; $p_{\rm tSZ} = 0.04$) achieve $p_{\rm tSZ} < 0.05$ in \texttt{SLOW}, though both still have large angular offsets from the observed clusters of $15.8^\circ$ and $13.4^\circ$, respectively; Virgo has $p_{\rm tSZ} = 0.09$ with an angular offset of $8.7^\circ$.
Notably, Coma---one of the most massive clusters in the local Universe---has $p_{\rm tSZ} = 0.87$ in \texttt{SLOW}, owing to the angular offset of ${\sim}15^\circ$ between the simulated halo and the observed counterpart, far larger than the ${\lesssim}1^\circ$ offsets typical of \ac{BORG}-based reconstructions (\cref{fig:Manti_clusters}).
By contrast, the majority of clusters in \CBM\ achieve $p_{\rm tSZ} < 0.05$, with the exceptions of Centaurus, Hydra, and Hercules (A2063), which have $\tilde{p}_{\rm tSZ} \approx 0.1$--$0.3$, and Hercules (A2151) and Abell~548, which have $\tilde{p}_{\rm tSZ} \approx 0.4$.
We also compute $p_{\rm LUM}$ for the \texttt{SLOW} clusters by evaluating the \ac{LUM} matching probability using the 3D separation between the \texttt{SLOW} halo positions and their observed counterparts (\cref{tab:slow_plum}).
Of the 14 matched \texttt{SLOW} clusters in~\cref{fig:cluster_ranking}, four achieve $p_{\rm LUM} < 0.05$: Virgo ($5.2 \times 10^{-4}$), Centaurus ($3.7 \times 10^{-3}$), Coma ($7.9 \times 10^{-3}$), and Shapley/A3558 ($4.5 \times 10^{-2}$).
The two metrics are complementary.
In \texttt{SLOW}, Coma achieves $p_{\rm LUM} = 0.008$ but $p_{\rm tSZ} = 0.87$ owing to its $14.9^\circ$ angular offset, while Perseus has $p_{\rm tSZ} = 0.008$ but $p_{\rm LUM} = 0.056$ due to a $21~\Mpch$ 3D separation.
Since \texttt{SLOW} provides a single realisation rather than a posterior ensemble, the $p_{\rm LUM}$ values are reported at face value without ensemble uncertainties, in contrast to the posterior spread shown for our simulations in \cref{tab:cluster_properties}.
For the majority of matched clusters, the ratio $p_{\rm LUM}^{\rm SLOW} / p_{\rm LUM}^{\rm CBM}$ exceeds two orders of magnitude (\cref{tab:slow_plum}), confirming that \ac{BORG}-based reconstructions achieve substantially better positioning.

\subsection{Scoring stacked profiles}\label{sec:scoring_stacked}

Rather than focusing on individual well-known systems, we now assess the mass dependence of the \ac{tSZ} signal by stacking one-dimensional radial profiles as a function of halo mass.
Following Section~\ref{sec:stacked_profiles}, we measure 1D radial profiles at normalised angular separations $\theta / (2\theta_{500{\rm c}})$, where $\theta_{500{\rm c}}$ is the angular radius corresponding to $R_{500{\rm c}}$, for haloes in each realisation and stack them in mass-ranked bins to capture the expected trend of increasing signal with mass.
We do not adjust the angular positions of the haloes; we simply extract the Compton-$y$ map within an aperture centred on each halo's sky position.
We define three cumulative bins: the top 10, top 50, and top 100 most massive haloes in each realisation, and then stack the resulting profiles across all realisations within each simulation.
This binning choice is designed to reveal the mass dependence while accounting for the systematic offset in halo masses between \CBb\ and \CBM\ (\cref{tab:cluster_properties}); ranking by mass within each realisation ensures a fair comparison despite this offset.

\begin{figure}
    \centering
    \includegraphics[width=\columnwidth]{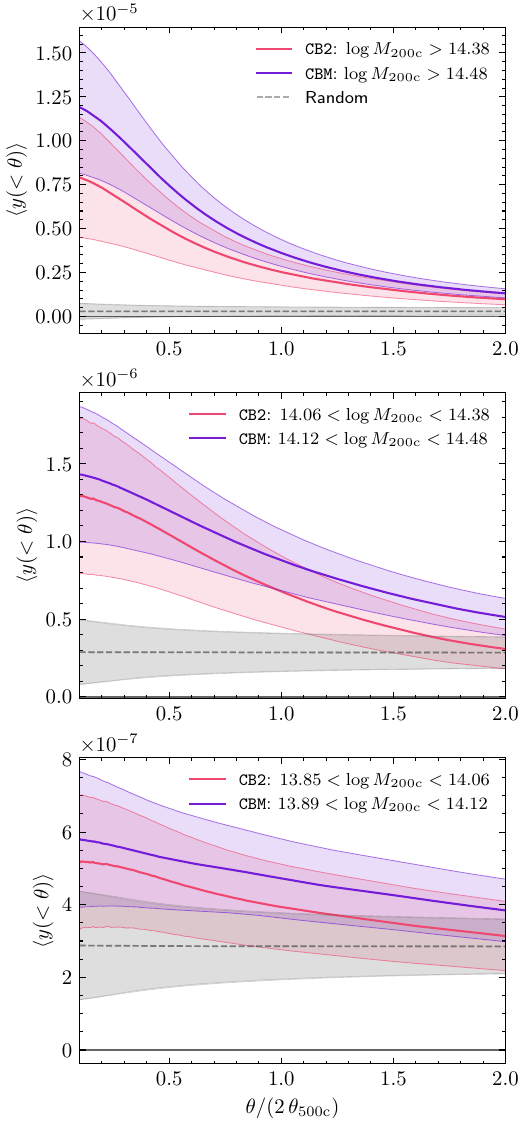}
    \caption{Stacked 1D radial \ac{tSZ} profiles for haloes ranked by mass. Profiles are measured at normalised angular separations $\theta / (2\theta_{500{\rm c}})$ and stacked within cumulative mass bins: top 10, top 50, and top 100 most massive haloes per realisation, then combined across all realisations. Halo masses $M_{200{\rm c}}$ are in units of $\Msunh$. Solid lines show the mean stacked profile for \CBb\ and \CBM, with shaded bands indicating the $1\sigma$ uncertainty from bootstrap resampling. The grey band shows the random expectation from stacking at random sky positions. Both simulations show a clear signal above random in the top 10 and top 50 bins, with \CBM\ marginally stronger in the top 10. The signal weakens in the top 100 bin, approaching the random baseline.}
    \label{fig:stacked_clusters}
\end{figure}

\Cref{fig:stacked_clusters} shows the stacked profiles for each bin, compared to random stacks constructed from uniformly distributed sky positions with aperture sizes drawn from the same distribution as the data.

For the top-10 and top-50 bins, both simulations yield a clear signal that is elevated above the random expectation across all radii.
In the top-10 bin, \CBM\ produces a marginally stronger signal than \CBb.
The top-50 bin shows a similar trend, with both simulations remaining well separated from the random baseline.
In the top-100 bin, the signal weakens and approaches the random expectation, primarily reflecting the fact that lower-mass haloes are not as well angularly aligned with their true counterparts.

\subsection{Scoring mass agreement}\label{sec:scoring_mass_agreement}

We now return to halo associations and match observed clusters using the classical matching procedure described in Section~\ref{sec:classical_matching}, requiring $\Delta\theta < 60^\prime$ and $\Delta cz < 300~\kmsec$.
The matched fractions are similar across simulations: for eROSITA~\citep{Bulbul2024_eRASS1}, we match 21/74 clusters in \CBb\ and 23/74 in \CBM; for \textit{Planck}, 13/56 in \CBb\ and 15/56 in \CBM.
For visualisation, we show the fraction of matched clusters as a function of the catalogue-reported mass in~\cref{fig:matching_fractions} (Appendix), dividing the catalogues into percentile bins above $M_{500{\rm c}} = 10^{14}~\Msun$; in the highest-mass bin, approximately half of the clusters are matched in both eROSITA and \textit{Planck} for both simulations.
These fractions would increase with looser matching criteria, but we opt for stringent thresholds to minimise spurious matches that would flatten the inferred scaling relation slopes.
This makes our results conservative.
Our matching criteria require precise agreement in both sky position ($\Delta\theta < 60^\prime$) and recession velocity ($\Delta cz < 300~\kmsec$), corresponding to spatial separations of no more than a few $\Mpch$ at $z < 0.05$.
This is a much more stringent standard than that adopted for \texttt{SLOW}~\citep{HernandezMartinez2024_SLOW}, where clusters separated by tens of $\Mpch$ can still be considered valid matches.

We exclude Shapley (A3558) from the mass scaling analysis due to a significant mass discrepancy: the \CBM\ halo has $\log (M_{500{\rm c}} / \Msun) = 15.24 \pm 0.12$, compared to $14.68 \pm 0.02$ from \textit{Planck} \ac{tSZ} and $14.81 \pm 0.04$ from eROSITA.
This may arise because A3558 lies near the edge of \TWOMPP\ completeness, where reconstruction fidelity is lower, or because \ac{BORG} lacks the resolution to separate neighbouring structures along the Shapley filament, instead merging them into a single massive halo.
The cluster itself is dynamically complex: dynamical mass estimates from~\citet{Haines2018_ShapleyFilaments} yield $M_{\rm dyn} = (14.8 \pm 1.4) \times 10^{14}~\Msun$, whereas the weak-lensing mass from~\citet{Higuchi_2020} is $\log (M_{200{\rm c}} / \Msunh) = 14.65^{+0.21}_{-0.33}$---a factor of ${\sim}2$ lower.
X-ray observations reveal large-scale sloshing and cold fronts extending to ${\sim}1.2~\rm Mpc$~\citep{Rossetti2007_A3558, Mirakhor2023_A3558}, consistent with A3558 being a post-merger system, possibly from a collision within the Shapley core~\citep{Bardelli1998_A3558}.

\subsubsection{tSZ--mass scaling}

The integrated Compton parameter $Y_{500{\rm c}}^{\rm tSZ}$ scales with cluster mass as $Y_{500{\rm c}}^{\rm tSZ} \propto M_{500{\rm c}}^m$, where self-similar theory predicts $m = 5/3$~\citep{Kaiser_1986}.
Observationally, the $Y_{500{\rm c}}^{\rm tSZ}$--$M$ relation is found to be remarkably close to this prediction, with slopes $m \approx 1.7$--$1.8$~\citep{Arnaud_2010,Planck_2014_SZcounts}.
For the matched \textit{Planck} clusters, we find slopes of $m = 1.43 \pm 0.31$ for \CBb\ and $m = 1.79 \pm 0.28$ for \CBM\ (\cref{tab:scaling_relations}, Appendix), both consistent with the self-similar expectation within uncertainties, though \CBM\ yields a slope closer to the expected value.
The Spearman correlation coefficients are high for both simulations ($\rho_{\rm s} = 0.74 \pm 0.11$), indicating a strong monotonic relationship between the \ac{BORG} halo masses and the observed $Y_{500{\rm c}}^{\rm tSZ}$.
\CBM\ yields a notably lower intrinsic scatter ($\sigma_{\rm int} = 0.09 \pm 0.07~\dex$) than \CBb\ ($\sigma_{\rm int} = 0.23 \pm 0.09~\dex$).

\subsubsection{tSZ and X-ray derived mass scaling}

We now examine the overall normalisation of the halo masses by comparing the \textit{Planck}-calibrated masses $M_{500{\rm c}}^{\rm tSZ}$ directly to the \ac{BORG} halo masses, where the expectation is a one-to-one relation ($m = 1$, $c = 0$).
For \CBM, we find a slope of $m = 1.12 \pm 0.17$, consistent with unity, whereas \CBb\ yields a shallower slope of $m = 0.76 \pm 0.18$.
The Spearman correlations are high for both simulations ($\rho_{\rm s} = 0.79 \pm 0.09$ for \CBM\ and $\rho_{\rm s} = 0.74 \pm 0.09$ for \CBb), indicating a strong monotonic relationship.
The intercepts indicate a systematic offset in mass normalisation: \CBM\ has $c = -0.20 \pm 0.07$, indicating that the \ac{BORG} halo masses are systematically larger than the \ac{tSZ}-calibrated masses by ${\sim}0.2~\dex$ (a factor of ${\sim}1.6$), with a joint tension of $4.7\sigma$ from the one-to-one relation.
For \CBb, the offset is weaker ($c = -0.02 \pm 0.07$, $2.2\sigma$ tension), though even here the majority of matched clusters have \ac{BORG} masses exceeding the \ac{tSZ} calibration.
However, these offsets are consistent with the known ${\sim}30\%$ bias of \textit{Planck} \ac{tSZ} masses relative to weak lensing calibrations~\citep{Sereno_2017}, suggesting that the \ac{BORG} halo masses may more accurately reflect the true cluster masses.

\begin{figure}
    \centering
    \includegraphics[width=\columnwidth]{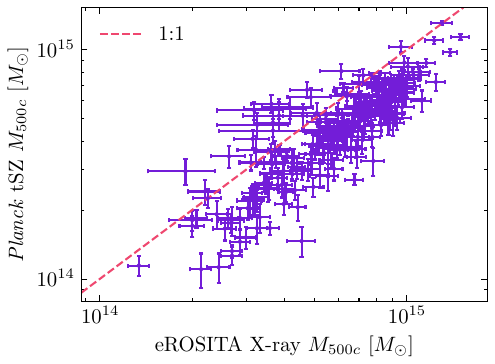}
    \caption{Comparison of Planck \ac{tSZ} masses and eROSITA X-ray masses for matched clusters; error bars show $1\sigma$ uncertainties. The dashed line shows the one-to-one relation. The Planck masses are systematically lower by ${\sim} 0.14~\dex$ (${\sim}30\%$), consistent with the known hydrostatic mass bias~\citep{Sereno_2017}.}
    \label{fig:mass_tSZ_eROSITA}
\end{figure}

We verify this bias by directly comparing the \textit{Planck} \ac{tSZ} masses to the weak-lensing-calibrated eROSITA X-ray masses for clusters matched between the two catalogues using $\Delta\theta < 15^\prime$ and $\Delta cz < 300~\kmsec$ (\cref{fig:mass_tSZ_eROSITA}).
We infer a linear relation in log-space using the \texttt{roxy}\footnote{\url{https://github.com/DeaglanBartlett/roxy}} package~\citep{Bartlett_2023}, which implements Marginalised Normal Regression (MNR) to account for measurement uncertainties in both variables by placing a Gaussian hyperprior on the true value of the independent variable, pivoting both variables at $\log (M_{500{\rm c}} / \Msun) = 14$.
We find a slope of $m = 1.01 \pm 0.04$, consistent with unity, and an intercept of $c = -0.14 \pm 0.03$, confirming that the \textit{Planck} \ac{tSZ} masses are systematically lower than the eROSITA X-ray masses by ${\sim} 0.14~\dex$ (${\sim}30\%$).
Given this known bias in the \ac{tSZ} mass calibration, it is not surprising that the \CBM\ halo masses---which are derived from the underlying dark matter distribution---exceed the \textit{Planck}-reported values.

\begin{figure*}
    \centering
    \begin{minipage}[b]{0.48\textwidth}
        \centering
        \includegraphics[width=\textwidth]{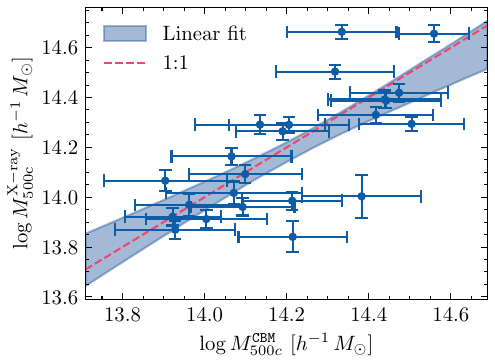}
        \\[0.5em]{\small (a) \CBM: $(m, c) = (0.84 \pm 0.19, 0.02 \pm 0.06)$, $\rho_{\rm s} = 0.62 \pm 0.10$}
    \end{minipage}
    \hfill
    \begin{minipage}[b]{0.48\textwidth}
        \centering
        \includegraphics[width=\textwidth]{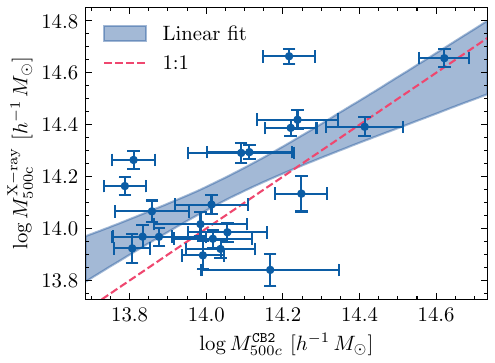}
        \\[0.5em]{\small (b) \CBb: $(m, c) = (0.75 \pm 0.21, 0.12 \pm 0.05)$, $\rho_{\rm s} = 0.49 \pm 0.10$}
    \end{minipage}
    \\[1em]
    \caption{eROSITA mass--mass relations comparing $M_{500{\rm c}}^{\rm X-ray}$ to the \ac{BORG} halo mass.
    The one-to-one relation is shown as a red dashed line.}
    \label{fig:erass_mass}
\end{figure*}

We perform an analogous comparison using the weak-lensing-calibrated eROSITA X-ray masses (\cref{fig:erass_mass}).
For \CBM, we find a slope of $m = 0.84 \pm 0.19$ and an intercept of $c = 0.02 \pm 0.06$, in excellent agreement with the one-to-one relation ($0.5\sigma$ tension), with a Spearman correlation of $\rho_{\rm s} = 0.62 \pm 0.10$.
In contrast, \CBb\ yields a shallower slope of $m = 0.75 \pm 0.21$ with a positive intercept $c = 0.12 \pm 0.05$, deviating from the expected relation at $2.1\sigma$; the correlation is also weaker ($\rho_{\rm s} = 0.49 \pm 0.10$).
This suggests that \CBM\ halo masses are well calibrated against the weak-lensing-based eROSITA masses, whereas \CBb\ masses are systematically lower at the high-mass end.

\begin{figure*}
    \centering
    \begin{minipage}[b]{0.48\textwidth}
        \centering
        \includegraphics[width=\textwidth]{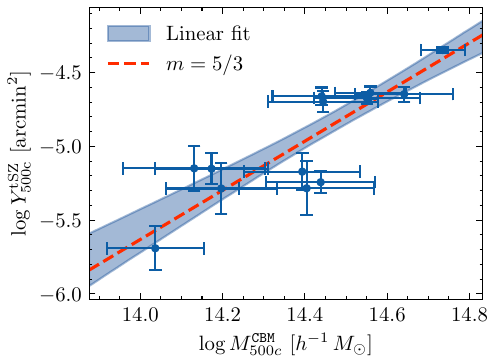}
        \\[0.5em]{\small (a) \CBM: $m = 1.79 \pm 0.28$, $\rho_{\rm s} = 0.74 \pm 0.10$}
    \end{minipage}
    \hfill
    \begin{minipage}[b]{0.48\textwidth}
        \centering
        \includegraphics[width=\textwidth]{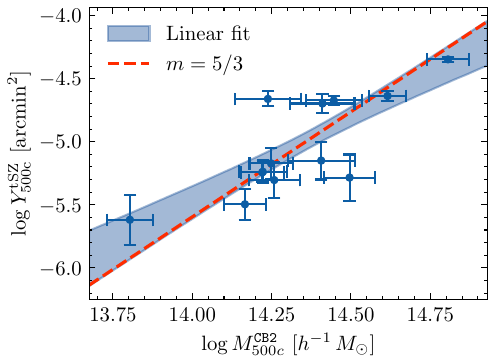}
        \\[0.5em]{\small (b) \CBb: $m = 1.43 \pm 0.31$, $\rho_{\rm s} = 0.74 \pm 0.11$}
    \end{minipage}
    \\[0.5em]
    \begin{minipage}[b]{0.48\textwidth}
        \centering
        \includegraphics[width=\textwidth]{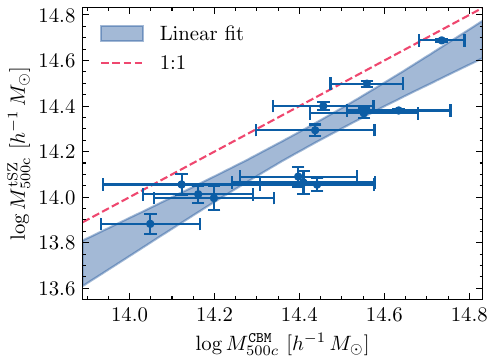}
        \\[0.5em]{\small (c) \CBM: $(m, c) = (1.12 \pm 0.17, -0.21 \pm 0.08)$, $\rho_{\rm s} = 0.79 \pm 0.09$}
    \end{minipage}
    \hfill
    \begin{minipage}[b]{0.48\textwidth}
        \centering
        \includegraphics[width=\textwidth]{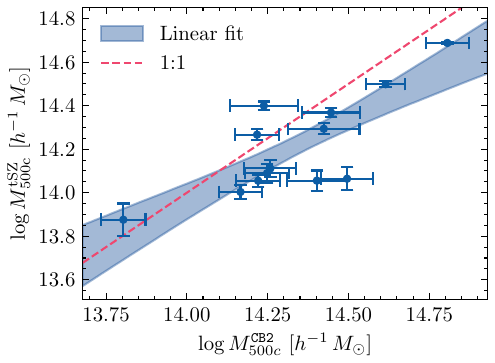}
        \\[0.5em]{\small (d) \CBb: $(m, c) = (0.76 \pm 0.18, -0.02 \pm 0.07)$, $\rho_{\rm s} = 0.74 \pm 0.09$}
    \end{minipage}
    \\[1em]
    \caption{Planck cluster scaling relations.
    \textit{Top row}: integrated Compton parameter $Y_{500{\rm c}}^{\rm tSZ}$ versus \ac{BORG} halo mass; the red dashed line shows the expected self-similar slope of $5/3$.
    \textit{Bottom row}: Planck-calibrated mass $M_{500{\rm c}}^{\rm tSZ}$ versus \ac{BORG} halo mass; the red dashed line shows the one-to-one relation.
    }
    \label{fig:planck_scaling}
\end{figure*}

\section{Discussion}\label{sec:discussion}

Bayesian reconstructions of the local Universe have advanced from early smoothed density field estimates to full digital twins that reproduce the positions, masses, and velocities of individual structures~\citep{Jasche_2019,Stopyra_2023,McAlpine_2025}.
As these reconstructions achieve ever-higher fidelity, robust validation demands an expanding suite of independent tests; velocity field comparisons offer one such metric~\citep{Stiskalek_2025_VFO}, but a broader range of observables is needed.
\citet{McAlpine_2025} proposed a comprehensive validation framework for \texttt{Manticore-Local}, including posterior predictive tests of power spectra, bispectra, the initial white noise field, and the halo mass function, alongside cluster matching and peculiar velocity comparisons.

Among such tests, the \ac{tSZ} effect offers a compelling probe: it traces the thermal energy of the intracluster medium, depends sensitively on halo mass through well-understood scaling relations, probes the angular positioning of clusters on the sky, and is measured with near-full-sky coverage by \textit{Planck}.
In this section, we first compare our results to other constrained simulation efforts (Section~\ref{sec:comparison_to_literature}), then discuss the validation metrics employed in this work (Section~\ref{sec:validation_metrics}), and finally discuss ways in which the work could be taken further (Section~\ref{sec:future}).

\subsection{Comparison to literature}\label{sec:comparison_to_literature}

Beyond the \ac{BORG} framework, constrained simulations of the local Universe have a long history of connecting observed large-scale structure to physically motivated predictions for cluster-scale observables.
Early work used IRAS galaxy data to constrain initial conditions within ${\sim}110~\Mpch$~\citep{Dolag2005_CORUSCANT}, enabling detailed comparisons between simulated and observed nearby clusters such as Coma and Virgo.
A particularly direct link to baryonic observables was established when the \textit{Planck} collaboration combined \ac{tSZ} and X-ray data with a constrained simulation of the Virgo environment to infer its distance and gas content~\citep{Planck2016_Virgo}, demonstrating that constrained realisations can successfully reproduce both the location and thermodynamic imprint of local clusters in Compton-$y$ maps.

Subsequent efforts refined these approaches by incorporating peculiar velocity data to better constrain the large-scale gravitational field.
The \texttt{CLUES} project~\citep{Gottloeber2010_CLUES,Sorce2016_CLUES} used distance indicators from the CosmicFlows programme to reduce cosmic variance on scales up to ${\sim}150~\Mpch$, providing improved predictions for the spatial distribution of massive haloes and their associated gas.
Building on this, the \texttt{SLOW} collaboration presented a $500~\Mpch$ constrained hydrodynamical simulation based on \texttt{CLUES} initial conditions and CosmicFlows-2 velocities~\citep{Dolag2023_SLOW,Tully2013_CF2}.
Using this framework, several studies investigated the thermodynamic properties of local clusters, including comparisons of X-ray and \ac{tSZ}-derived masses~\citep{HernandezMartinez2024_SLOW}, thermodynamic profiles and evolution of local clusters~\citep{HernandezMartinez2025_SLOW}, projection effects in Virgo~\citep{Sorce2021_CLONE,Lebeau2024_VirgoMassBias}, and the contribution of local \ac{tSZ} and \ac{kSZ} signals to large-scale \ac{CMB} anomalies~\citep{Jung2024_CMBanomalies}.

These velocity-based constrained simulations provide a complementary perspective to \ac{BORG}-based reconstructions, which infer initial conditions from galaxy redshift surveys such as \TWOMPP\ using a fully Bayesian forward model.
An example of this approach applied to galaxy observables is the \texttt{SIBELIUS-DARK} simulation~\citep{McAlpine2022_SIBELIUS}, which produced a constrained realisation of the local volume within ${\sim}200~\Mpch$ based on \ac{BORG} initial conditions from~\citet{Jasche_2019}, with additional constraints on the small-scale modes not captured by \ac{BORG} so as to reproduce the Local Group, evolved with the \textsc{eagle} galaxy formation model~\citep{Schaye2015_EAGLE}, enabling explicit cluster matching against observed counterparts.
Because the two approaches rely on different data and assumptions, cross-validation using observables such as the \ac{tSZ} signal is essential.
Indeed, \citet{Stiskalek_2025_VFO} showed that \ac{BORG}-based reconstructions reproduce peculiar velocity samples more accurately than \texttt{CLUES}-based initial conditions~\citep{Sorce2018_CLUES,Sorce2020_CLUES}, implying a more faithful reconstruction of the local velocity field.
This improvement is reflected in cluster positions: while \texttt{SLOW} exhibits typical offsets of ${\gtrsim}10~\Mpch$ for well-constrained systems~\citep{HernandezMartinez2024_SLOW}, the matching criteria adopted here correspond to spatial separations of only a few $\Mpch$ at low redshift, highlighting the tighter angular and positional agreement achieved by \ac{BORG}-based digital twins such as \CBM\ (see \cref{tab:cluster_properties} for the separations in \CBM).
Our \ac{tSZ}-based validation confirms this quantitatively: of the clusters in our sample (\cref{tab:named_clusters}), only two \texttt{SLOW} counterparts achieve $p_{\rm tSZ} < 0.05$, whereas the majority of clusters in \CBb\ and \CBM\ are detected at much higher significance (\cref{fig:cluster_ranking}).
A comparison using $p_{\rm LUM}$, which accounts for 3D separation and halo mass, yields four \texttt{SLOW} clusters with $p_{\rm LUM} < 0.05$---all among the most massive and nearest systems (\cref{tab:slow_plum})---whereas the \ac{BORG}-based simulations achieve $p_{\rm LUM} < 0.05$ for the majority of their matched clusters, with $p_{\rm LUM}$ ratios typically exceeding $10^2$ (\cref{tab:cluster_properties,tab:slow_plum}).
A more detailed $p_{\rm LUM}$ comparison between Manticore and a peculiar-velocity-based reconstruction is presented in \citet{McAlpine2025_ManticoreCatalogue}.

\subsection{Validation metrics}\label{sec:validation_metrics}

Our approach to quantifying the quality of the match between simulated clusters and the \ac{tSZ} signal complements the \ac{LUM} framework of~\citet{LUM}, which scores reconstructions by comparing simulated and observed cluster positions in three-dimensional redshift space to associate simulated haloes with observed counterparts.
While powerful, this method is complicated by uncertainties in cluster distances, which can exceed several hundred $\kmsec$ even for nearby systems.
By contrast, our \ac{tSZ}-based test focuses on angular agreement with observed hotspots, sidestepping radial distance uncertainties entirely; the goal here is not to establish halo--cluster associations but simply to assess significance against the \ac{tSZ} map.
This enables the derivation of per-cluster significance values that quantify how well the reconstructed halo positions coincide with the thermal gas distribution on the sky.

\begin{figure}
    \centering
    \includegraphics[width=\columnwidth]{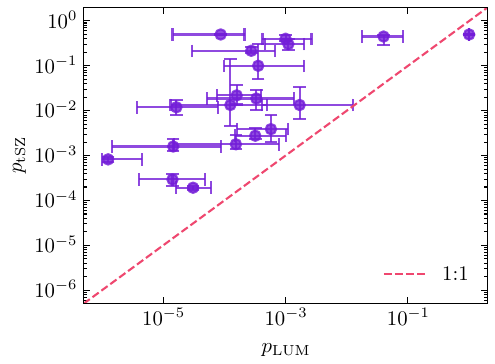}
    \caption{Comparison of $p_{\rm LUM}$-values and $p_{\rm tSZ}$-values for selected clusters (\cref{tab:named_clusters}) in \CBM. Points show the median $p$-value across realisations for each cluster, with error bars indicating the 16th--84th percentile spread. The red dashed line indicates 1:1 correspondence. The two metrics are correlated, though the \ac{tSZ} test is generally stricter.}
    \label{fig:lum_vs_tsz}
\end{figure}

Because $p_{\rm tSZ}$ is based primarily on angular position, with only a secondary dependence on the aperture size set by the halo's physical extent, a halo placed at the correct $(\ell, b)$ but at an incorrect distance would still score well, and multiple structures along the same line of sight contribute to the projected Compton-$y$ signal.
In practice, these effects are minor for our sample: at $z < 0.05$ the Compton-$y$ signal is dominated by the target cluster, making chance alignments with comparably massive systems unlikely, though line-of-sight contamination would need to be accounted for at higher redshifts.
Moreover, for the named-cluster sample we also report $p_{\rm LUM}$-values (\cref{tab:cluster_properties}), which incorporate three-dimensional distance and mass information, and compare the two metrics in \cref{fig:lum_vs_tsz}.

\Cref{fig:lum_vs_tsz} compares $p_{\rm LUM}$-values, computed from the probability that a halo at a given distance could produce the observed cluster position given its mass, with $p_{\rm tSZ}$-values for selected clusters.
The two metrics are weakly correlated, as expected since both probe positional agreement, though the \ac{tSZ} test is generally stricter, yielding lower $p_{\rm tSZ}$-values for clusters that are marginally significant under the \ac{LUM} criterion.
The large spread in $p_{\rm LUM}$-values for individual clusters reflects the spatial extent of halo associations: haloes near the observed position have low $p_{\rm LUM}$-values, while more distant members have high $p_{\rm LUM}$-values.

The interpretation of $p_{\rm tSZ}$ depends on the data used to constrain the initial conditions.
\ac{BORG} infers initial conditions from the \TWOMPP\ galaxy number density field, which encodes the positions of galaxy overdensities on a coarse $3.9~\Mpc$ grid; although cluster positions must still emerge through non-linear gravitational evolution in the $N$-body simulation, the angular coincidence with \ac{tSZ} hotspots may be expected to be easier to achieve with this input.
However, the \TWOMPP\ catalogue used in \ac{BORG} does not apply any group finding, so cluster cores remain affected by redshift-space distortions that smear their apparent positions along the line of sight, complicating the inference of precise cluster locations.
Peculiar-velocity-based reconstructions such as \texttt{SLOW} use distance indicators---such as the Tully--Fisher relation~\citep{TullyFisher1977}---that carry angular position information but whose distances are inferred with typical uncertainties of up to ${\sim}25\%$, whereas the \ac{BORG} redshift-space likelihood is dominated by Poisson noise in the galaxy counts, which yields better reconstructions even when tested against peculiar velocity data~\citep{Stiskalek_2025_VFO}.
Since even catalogues such as CosmicFlows-4~\citep{Tully2023_CF4} track the sky positions of cluster-hosting overdensities, the angular fidelity of the output is also a test of the forward model, not only a reflection of the observational inputs.
However, despite these differences, one key science goal of constrained reconstructions is the prediction of \ac{CMB} secondary anisotropies---e.g.\ \ac{tSZ}, \ac{kSZ}, and gravitational lensing---which requires accurate sky positions of massive structures.

Second, we derive stacked \ac{tSZ} profiles centred on simulated haloes, similarly to~\citet{McAlpine2025_ManticoreCatalogue}, and compare two generations of \ac{BORG} reconstructions.
While sensitive to angular agreement like the per-cluster test, stacking addresses how well structures are reconstructed as a function of halo mass on average.

Third, we introduce a metric that goes beyond angular agreement to test the scaling between halo mass and the observable of an associated cluster.
Even on the \ac{tSZ} side, extracting the integrated Compton parameter relies on assumptions about the cluster extent, introducing model dependence.
Direct mass comparisons are more complicated still: both \ac{tSZ} and X-ray masses rely on external calibrations---hydrostatic equilibrium or weak lensing---each with its own systematics.
Of these, weak-lensing calibration is generally more reliable, as it measures the total projected mass without assuming dynamical equilibrium.

Part of the purpose here is to build larger matched samples over which to test the mass calibration of individual clusters.
This must be balanced against spurious matches, for which reason we adopt strict matching criteria of $\Delta\theta < 60^\prime$ and $\Delta cz < 300~\kmsec$ (Section~\ref{sec:classical_matching}).
We have verified that under these criteria very few haloes from the earlier~\citet{Jasche_2019} initial conditions in the \CBa\ suite~\citep{Bartlett2021_CSiBORG} would be matched to observed \ac{tSZ} counterparts, underscoring the improvement in reconstruction fidelity achieved by later generations of \texttt{BORG}.

We have not yet formalised our mass-based metrics into a single Bayesian evidence; instead, we examine the inferred slope and intercept of power-law scaling relations, the intrinsic scatter, and the Spearman correlation coefficient.
A proper quantification would compute the Bayesian evidence, as done for velocity field comparisons by~\citet{Stiskalek_2025_VFO}.
The evidence depends on both the model and the data: in the limit of independent data points, the likelihood factorises over individual cluster contributions, and the evidence is the integral of this likelihood weighted by the prior over model parameters.
However, a complication arises because different simulations yield different numbers of matched clusters---for instance, one reconstruction may fail to reproduce a particular cluster entirely.
A straightforward evidence comparison would therefore be ambiguous unless one also penalises local Universe models that fail to reproduce observed systems.
Developing a fully probabilistic matching framework that naturally incorporates such penalties, enabling rigorous Bayesian model comparison across reconstructions, is a key direction for future work.

\subsection{Future directions}\label{sec:future}

Digital twins provide a complementary framework to purely statistical analyses of the \ac{tSZ} signal.
While standard \ac{tSZ} studies extract cosmological information from ensemble statistics of the Compton-$y$ field, digital twins retain the spatial phase information of the local density and velocity fields, enabling object-level comparisons and physically motivated interpretations.
By anchoring the \ac{tSZ} signal to a posterior set of large-scale structure realisations, digital twins allow explicit modelling of projection effects, cluster environments, and line-of-sight contributions that are otherwise absorbed into effective scatter or noise terms.
This approach is particularly valuable in the local Universe, where cosmic variance and individual massive clusters dominate the \ac{tSZ} signal, and demonstrates how digital twins can augment statistical \ac{tSZ} analyses by providing additional physical context.
In our case, the context is ``learnt'' from large-scale structure (the galaxy number density field) by the \texttt{BORG} algorithm under the assumption of the standard cosmological model.

Besides developing a full evidence comparison between different reconstructions as described above, we note three ways in which this work could be taken forward.
First, incorporating \ac{tSZ} or X-ray cluster data as additional likelihood terms within the \texttt{BORG} framework would improve constraints on halo masses and positions in under-sampled regions where galaxy redshift surveys provide limited information.
This would improve the precision of the inferred initial conditions of the local Universe, further improving the fidelity of the reconstructions.
A key technical challenge, however, is forward-modelling cluster masses within a differentiable inference pipeline, as halo identification and mass estimation are not straightforwardly differentiable operations.
Recent work has demonstrated gradient-free optimisation of initial conditions using non-differentiable structure formation models that can be extended efficiently to larger volumes~\citep{Doeser2025_LULO}, though extending this to full posterior sampling remains an open problem.
Alternatively, differentiable halo finders have been developed~\citep{Horowitz2025_jFoF}, though discrete particle membership remains a fundamental obstacle to stable gradient propagation.

The second extension would be to use \texttt{BORG} in conjunction with the \ac{CMB} to obtain new constraints on astrophysical objects.
The approach here would be to forward-model Compton-$y$ (and/or X-ray) maps using \CBM\ cluster catalogues plus a scheme for ``baryonifying'' the dark matter structures by using semi-empirical or semi-analytic models for the free electron distribution, temperature and pressure profile (e.g.~\citealt{Arnaud_2010,Battaglia_1,Battaglia_2}).
Comparing to the measurements with a pixel-by-pixel likelihood would then afford precise constraints on these baryonic parameters.
This is similar to previous work constraining baryons with the \ac{tSZ} signal~\citep{Koukoufilippas2020,Yan2021,Troster2022,Koukoufilippas2022,LaPosta2025}, but crucially the use of constrained simulations provides the connection to dark matter on an object-by-object basis.

The final stage would be to use \texttt{BORG} to assist in modelling the \ac{CMB} itself.
The idea here would be to forward-model all relevant secondary anisotropies using \texttt{BORG} posterior fields to the map level---i.e. $I_\nu(n)$ (plus polarisation in principle)---and then analyse the \ac{CMB} data jointly with the \ac{CMB} primary, foregrounds (which could also be informed by \texttt{BORG}) and instrumental effects.
This would integrate the study of large-scale structure with the study of the \ac{CMB}, allowing inferences from both to be made simultaneously and explicitly self-consistently.

\section{Conclusion}\label{sec:conclusion}

We have validated two generations of \ac{BORG}-based constrained simulations of the local Universe---\CBb\ (initial conditions from~\citealt{Stopyra_2023}) and \CBM\ (initial conditions from~\citealt{McAlpine_2025})---against \textit{Planck} \ac{tSZ} and eROSITA X-ray cluster observations.
To do so, we introduced three complementary tests: per-cluster angular positioning using $p_{\rm tSZ}$-values derived from the Compton-$y$ map, stacked radial profiles as a function of halo mass, and mass scaling relations for matched cluster samples.

The majority of selected clusters are reproduced with high fidelity ($p_{\rm tSZ} < 0.05$), with Coma, Shapley (A3558) and Perseus among the best-constrained systems.
By contrast, only two clusters in \texttt{SLOW}~\citep{HernandezMartinez2024_SLOW}---a hydrodynamical resimulation of \texttt{CLUES} velocity-constrained initial conditions~\citep{Sorce2016_CLUES}---achieve $p_{\rm tSZ} < 0.05$, while four achieve $p_{\rm LUM} < 0.05$ (\cref{tab:slow_plum}), underscoring the superior positional accuracy of \ac{BORG}-based digital twins.
Perseus is absent in \CBb~\citep{Stopyra_2023} but well reproduced in \CBM~\citep{McAlpine_2025}.
\CBM\ yields systematically higher halo masses than \CBb, in good agreement with weak-lensing-calibrated eROSITA masses as shown by~\citet{McAlpine2025_ManticoreCatalogue}.
This demonstrates that future field-level models of the \ac{CMB} informed by large-scale structure should use \ac{BORG}---specifically its latest incarnation in \texttt{Manticore}---as their foundation.

The \ac{tSZ} effect provides a powerful validation observable that complements velocity field comparisons, probing the angular positioning and masses of clusters independently of distance uncertainties.
By matching simulated haloes to observed clusters, digital twins can be used to study mass calibration on a per-cluster basis, offering an alternative to stacked weak lensing analyses.

Looking ahead, formalising these metrics into a Bayesian evidence framework would enable rigorous model comparison across reconstructions, while incorporating \ac{tSZ} and X-ray cluster data as likelihood terms directly into the \ac{BORG} inference could further improve constraints on halo masses and positions in regions where galaxy redshift surveys provide limited information.
Our work also paves the way for inferring astrophysical parameters by forward-modelling Compton-$y$ maps using \texttt{BORG}-based halo catalogues, and even more ambitiously for inferring large-scale structure and the \ac{CMB} (primary, secondary and foregrounds) simultaneously and self-consistently using an integrated field-level likelihood.

\section*{Acknowledgements}

We thank David Alonso, Mayeul Aubin, Pedro Ferreira, Jens Jasche, and Guilhem Lavaux for useful inputs and discussions.
RS acknowledges financial support from STFC Grant No. ST/X508664/1 and the Snell Exhibition of Balliol College, Oxford.
HD is supported by a Royal Society University Research Fellowship (grant no. 211046).
This work was done within the Aquila Consortium\footnote{\url{https://aquila-consortium.org}}.

\section*{Data availability}

The code used in this work is available at \url{https://github.com/Richard-Sti/CMBOlympics}.
The \texttt{Manticore} data products are available at \url{https://digitaltwin.fysik.su.se}.
The eROSITA DR1 cluster catalogue is available at \url{https://erosita.mpe.mpg.de/dr1/AllSkySurveyData_dr1/Catalogues_dr1/}.
The \textit{Planck} PSZ2 cluster catalogue is available at \url{https://heasarc.gsfc.nasa.gov/w3browse/all/plancksz2.html}.
The \textit{Planck} Compton-$y$ map was publicly released by~\citet{McCarthyHill2024}.
All other data are available upon reasonable request.

\bibliographystyle{apsrev4-1}
\bibliography{ref}

\appendix

\section{Scaling relation inference}\label{app:scaling_relations}

\Cref{tab:scaling_relations} summarises the inferred scaling relation parameters for both observable--mass and mass--mass relations.
We infer relations of the form $\log Y = m \log M + c$ with intrinsic scatter $\sigma_{\rm int}$, where $Y$ is either the catalogue observable or catalogue mass, and $M$ is the matched \ac{BORG} halo mass.
The Spearman correlation coefficient $\rho_{\rm s}$ quantifies the strength of the correlation, while $\tau_m$ and $\tau_{m,c}$ give the tension with the self-similar slope ($m = 4/3$ for X-ray luminosity, $m = 5/3$ for \ac{tSZ};~\citealt{Kaiser_1986}) and unity relation ($m = 1$, $c = 0$), respectively.

\begin{table*}
    \centering
    \renewcommand{\arraystretch}{1.3}
    \begin{tabular}{llccccc}
        \hline
        \multicolumn{7}{c}{\textbf{Observable--mass scaling relations}} \\
        \hline
        Catalogue & Simulation & $m$ & $c$ & $\sigma_{\rm int}$ & $\rho_{\rm s}$ & $\tau_m$ \\
        \hline
        eROSITA & \CBb   & $1.39 \pm 0.34$ & $1.46 \pm 0.08$ & $0.30 \pm 0.06$ & $0.51 \pm 0.09$ & $0.2\sigma$ \\
                & \CBM & $1.63 \pm 0.34$ & $1.26 \pm 0.10$ & $0.21 \pm 0.08$ & $0.63 \pm 0.10$ & $0.9\sigma$ \\
        \hline
        Planck & \CBb   & $1.43 \pm 0.31$ & $-1.51 \pm 0.13$ & $0.22 \pm 0.10$ & $0.74 \pm 0.11$ & $0.8\sigma$ \\
               & \CBM & $1.79 \pm 0.28$ & $-1.73 \pm 0.15$ & $0.08 \pm 0.07$ & $0.74 \pm 0.10$ & $0.4\sigma$ \\
        \hline
        \\[-0.5em]
        \hline
        \multicolumn{7}{c}{\textbf{Mass--mass relations}} \\
        \hline
        Catalogue & Simulation & $m$ & $c$ & $\sigma_{\rm int}$ & $\rho_{\rm s}$ & $\tau_{m,c}$ \\
        \hline
        eROSITA & \CBb   & $0.75 \pm 0.21$ & $0.12 \pm 0.05$ & $0.18 \pm 0.04$ & $0.49 \pm 0.10$ & $2.1\sigma$ \\
                & \CBM & $0.84 \pm 0.19$ & $0.02 \pm 0.06$ & $0.14 \pm 0.04$ & $0.62 \pm 0.10$ & $0.5\sigma$ \\
        \hline
        Planck & \CBb   & $0.76 \pm 0.18$ & $-0.02 \pm 0.07$ & $0.14 \pm 0.04$ & $0.74 \pm 0.09$ & $2.2\sigma$ \\
               & \CBM & $1.12 \pm 0.17$ & $-0.21 \pm 0.08$ & $0.05 \pm 0.04$ & $0.79 \pm 0.09$ & $4.7\sigma$ \\
        \hline
    \end{tabular}
    \caption{Scaling relation inference of the form $\log Y = m \log M + c$, where $M$ is in units of $10^{14}~\Msunh$. \emph{Top}: observable--mass relations, where $Y$ is $L_{500{\rm c}}^{\rm X-ray}$ in units of $10^{44}~{\rm erg\,s^{-1}}$ (eROSITA) or $Y_{500{\rm c}}^{\rm tSZ}$ in units of $10^{-4}~{\rm Mpc}^2$ (Planck); $\tau_m$ gives the tension with the self-similar expectation ($m = 4/3$ for X-ray, $m = 5/3$ for \ac{tSZ}). \emph{Bottom}: mass--mass relations comparing catalogue $M_{500{\rm c}}$ to \ac{BORG} halo masses, both in units of $10^{14}~\Msunh$; $\tau_{m,c}$ gives the joint tension with the one-to-one relation ($m = 1$, $c = 0$).}
    \label{tab:scaling_relations}
\end{table*}

\section{Cluster sample properties}\label{app:cluster_properties}

\begin{figure*}
    \centering
    \includegraphics[width=0.48\textwidth]{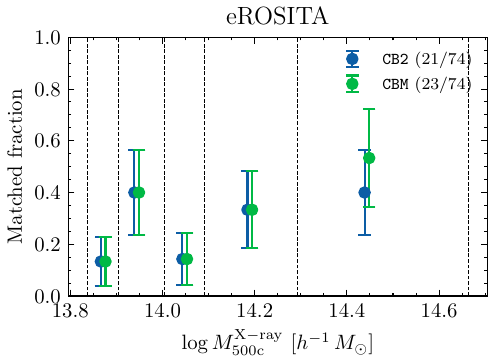}
    \includegraphics[width=0.48\textwidth]{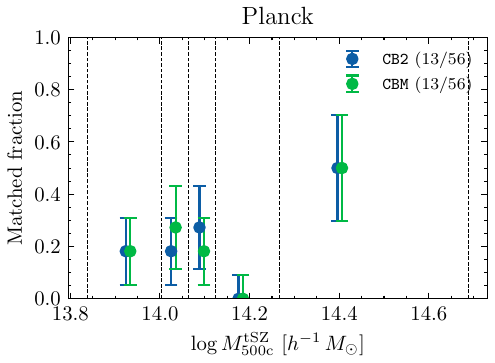}
    \caption{Fraction of observed clusters matched to halo associations as a function of catalogue-reported mass $M_{500{\rm c}}$ for eROSITA (left) and Planck (right). Clusters are divided into percentile bins above $M_{500{\rm c}} = 10^{14}~\Msunh$, with vertical dashed lines indicating bin edges. Error bars show Poisson uncertainties. The total matched fractions are indicated in the legend. Higher-mass clusters are more likely to be matched, with approximately half of the most massive clusters having counterparts in both simulations.}
    \label{fig:matching_fractions}
\end{figure*}

\Cref{fig:matching_fractions} shows the fraction of observed clusters matched to halo associations as a function of catalogue-reported mass; higher-mass clusters are more likely to have counterparts in both simulations.
\Cref{tab:cluster_properties} summarises the properties of the selected nearby clusters (Section~\ref{sec:named_clusters}) and their matched halo associations in \CBb\ and \CBM.

\vspace{1em}

\begin{table*}
    \centering
    \renewcommand{\arraystretch}{1.2}
    \begin{tabular}{lcccccccc}
        \hline
        \multicolumn{9}{c}{\textbf{\CBb}} \\
        \hline
        Cluster & $d$ [$h^{-1}\,{\rm Mpc}$] & 3D sep [$h^{-1}\,{\rm Mpc}$] & $\ell$ [deg] & $b$ [deg] & $f_{\rm present}$ & $\log M_{500{\rm c}}$ [$h^{-1}\,M_\odot$] & $p_{\rm LUM}$ & $\tilde{p}_{\rm tSZ}$ \\
        \hline
        Virgo             &  10.5 & 3.1 & 291.6 &  74.7 & 0.80 & 13.97 & $5.3 \times 10^{-3}$ & $4.7 \times 10^{-3}$ \\
        Centaurus (A3526) &  33.1 & 2.3 & 302.9 &  21.2 & 0.85 & 13.82 & $4.1 \times 10^{-3}$ & $1.2 \times 10^{-1}$ \\
        Hydra (A1060)     &  41.1 & 1.6 & 269.8 &  26.5 & 1.00 & 14.09 & $5.1 \times 10^{-4}$ & $2.3 \times 10^{-1}$ \\
        Norma (A3627)     &  48.8 & 4.1 & 325.5 &  $-7.5$ & 0.85 & 13.88 & $2.1 \times 10^{-2}$ & --- \\
        Perseus (A426)    &   --- & --- &   --- &   --- &  --- &   --- & --- & --- \\
        Leo (A1367)       &  63.3 & 1.6 & 234.2 &  73.3 & 1.00 & 14.22 & $3.2 \times 10^{-4}$ & $6.0 \times 10^{-3}$ \\
        Coma (A1656)      &  70.9 & 2.4 &  61.5 &  88.3 & 1.00 & 14.82 & $4.5 \times 10^{-5}$ & $2.4 \times 10^{-4}$ \\
        Hercules (A2199)  &  94.4 & 3.0 &  63.4 &  43.8 & 1.00 & 14.31 & $1.1 \times 10^{-3}$ & $5.6 \times 10^{-3}$ \\
        Abell 496         &  96.8 & 0.8 & 209.1 & $-36.6$ & 0.95 & 14.21 & $3.8 \times 10^{-5}$ & $2.0 \times 10^{-2}$ \\
        Hercules (A2063)  & 102.8 & 1.1 &  12.7 &  49.5 & 1.00 & 14.26 & $1.1 \times 10^{-4}$ & $3.8 \times 10^{-2}$ \\
        Hercules (A2151)  &   --- & --- &   --- &   --- &  --- &   --- & --- & --- \\
        Hercules (A2147)  & 108.2 & 3.6 &  30.6 &  44.3 & 1.00 & 14.55 & $7.6 \times 10^{-4}$ & $9.1 \times 10^{-3}$ \\
        Shapley (A3571)   & 114.3 & 1.4 & 316.5 &  28.7 & 0.95 & 14.61 & $3.1 \times 10^{-5}$ & $1.8 \times 10^{-3}$ \\
        Abell 548         & 120.3 & 4.5 & 230.2 & $-24.5$ & 1.00 & 14.08 & $1.1 \times 10^{-2}$ & $3.8 \times 10^{-1}$ \\
        Abell 119         & 125.8 & 2.4 & 125.5 & $-64.1$ & 1.00 & 14.44 & $4.5 \times 10^{-4}$ & $1.4 \times 10^{-3}$ \\
        Abell 1736        & 136.2 & 3.0 & 312.8 &  35.0 & 0.95 & 14.41 & $9.0 \times 10^{-4}$ & $5.2 \times 10^{-3}$ \\
        Abell 1644        & 139.2 & 0.5 & 305.0 &  45.3 & 1.00 & 14.25 & $1.4 \times 10^{-5}$ & $3.3 \times 10^{-3}$ \\
        Shapley (A3558)   & 144.0 & 0.8 & 312.3 &  30.6 & 1.00 & 14.93 & $6.4 \times 10^{-7}$ & $6.9 \times 10^{-4}$ \\
        \\[-0.5em]
        \hline
        \multicolumn{9}{c}{\textbf{\CBM}} \\
        \hline
        Cluster & $d$ [$h^{-1}\,{\rm Mpc}$] & 3D sep [$h^{-1}\,{\rm Mpc}$] & $\ell$ [deg] & $b$ [deg] & $f_{\rm present}$ & $\log M_{500{\rm c}}$ [$h^{-1}\,M_\odot$] & $p_{\rm LUM}$ & $\tilde{p}_{\rm tSZ}$ \\
        \hline
        Virgo             &  13.1 & 1.8 & 288.4 &  73.3 & 0.96 & 14.24 & $5.8 \times 10^{-4}$ & $3.9 \times 10^{-3}$ \\
        Centaurus (A3526) &  34.1 & 2.3 & 302.8 &  21.6 & 0.96 & 14.20 & $1.1 \times 10^{-3}$ & $3.0 \times 10^{-1}$ \\
        Hydra (A1060)     &  43.4 & 1.3 & 269.6 &  26.4 & 1.00 & 14.18 & $2.8 \times 10^{-4}$ & $2.2 \times 10^{-1}$ \\
        Norma (A3627)     &  51.2 & 2.0 & 325.3 &  $-6.3$ & 0.96 & 14.66 & $8.7 \times 10^{-5}$ & --- \\
        Perseus (A426)    &  51.2 & 1.5 & 150.3 & $-13.5$ & 1.00 & 14.75 & $1.4 \times 10^{-5}$ & $2.9 \times 10^{-4}$ \\
        Leo (A1367)       &  66.7 & 0.6 & 233.4 &  73.2 & 0.98 & 14.44 & $1.6 \times 10^{-5}$ & $1.2 \times 10^{-2}$ \\
        Coma (A1656)      &  72.2 & 1.8 &  62.3 &  88.2 & 1.00 & 14.74 & $3.1 \times 10^{-5}$ & $1.9 \times 10^{-4}$ \\
        Hercules (A2199)  &  94.3 & 3.0 &  63.4 &  43.7 & 1.00 & 14.68 & $1.6 \times 10^{-4}$ & $2.2 \times 10^{-2}$ \\
        Abell 496         & 100.7 & 2.0 & 209.4 & $-36.5$ & 0.88 & 14.37 & $3.3 \times 10^{-4}$ & $1.9 \times 10^{-2}$ \\
        Hercules (A2063)  & 100.6 & 1.6 &  12.3 &  49.5 & 0.96 & 14.28 & $3.6 \times 10^{-4}$ & $1.0 \times 10^{-1}$ \\
        Hercules (A2151)  & 118.5 & 5.8 &  32.1 &  44.2 & 0.54 & 13.94 & $4.0 \times 10^{-2}$ & $4.5 \times 10^{-1}$ \\
        Hercules (A2147)  & 106.0 & 3.3 &  29.9 &  44.4 & 0.94 & 14.63 & $3.2 \times 10^{-4}$ & $2.8 \times 10^{-3}$ \\
        Shapley (A3571)   & 113.6 & 0.8 & 316.5 &  28.6 & 1.00 & 14.57 & $1.5 \times 10^{-5}$ & $1.6 \times 10^{-3}$ \\
        Abell 548         & 117.8 & 3.9 & 230.4 & $-25.0$ & 0.90 & 14.52 & $1.0 \times 10^{-3}$ & $4.0 \times 10^{-1}$ \\
        Abell 119         & 125.3 & 2.1 & 125.5 & $-63.8$ & 0.96 & 14.58 & $1.5 \times 10^{-4}$ & $1.8 \times 10^{-3}$ \\
        Abell 1736        & 136.5 & 4.3 & 312.4 &  34.9 & 0.84 & 14.45 & $1.7 \times 10^{-3}$ & $1.3 \times 10^{-2}$ \\
        Abell 1644        & 140.5 & 1.0 & 305.0 &  45.3 & 0.72 & 14.46 & $1.3 \times 10^{-4}$ & $1.3 \times 10^{-2}$ \\
        Shapley (A3558)   & 145.5 & 1.4 & 312.2 &  30.6 & 0.90 & 15.07 & $1.3 \times 10^{-6}$ & $8.2 \times 10^{-4}$ \\
        \hline
    \end{tabular}
    \caption{Properties of the selected clusters and their matched halo associations for \CBb\ (top) and \CBM\ (bottom). We report the comoving distance $d$ to the matched halo association, 3D separation (3D sep) between the observed and simulated cluster positions, Galactic coordinates $(\ell, b)$, association presence fraction $f_{\rm present}$, mean logarithmic halo mass $\log M$, the \ac{LUM} $p$-value $p_{\rm LUM}$ quantifying match significance, and the median \ac{tSZ} $p$-value $\tilde{p}_{\rm tSZ}$ across realisations. Clusters without a matched association are marked with dashes.}
    \label{tab:cluster_properties}
\end{table*}

\begin{table*}
    \centering
    \renewcommand{\arraystretch}{1.15}
    \begin{tabular}{lccccccc}
        \hline
        Cluster & $\Delta\theta$ [deg] & $\Delta r$ [$\Mpch$] & $\log M_{500}$ [$M_\odot$] & $p_{\rm LUM}^{\rm SLOW}$ & $p_{\rm tSZ}^{\rm SLOW}$ & $p_{\rm LUM}^{\rm SLOW} / p_{\rm LUM}^{\rm CBM}$ & $p_{\rm tSZ}^{\rm SLOW} / \tilde{p}_{\rm tSZ}^{\rm CBM}$ \\
        \hline
        Virgo             &   8.7 &   5.6 & 14.82 & $5.2 \times 10^{-4}$ & $9.3 \times 10^{-2}$ & $0.9$       & $24$ \\
        Centaurus (A3526) &  16.3 &  10.3 & 14.80 & $3.7 \times 10^{-3}$ & $2.9 \times 10^{-1}$ & $3$         & $1$ \\
        Hydra (A1060)     &  12.4 &  12.6 & 14.42 & $6.3 \times 10^{-2}$ & $9.6 \times 10^{-1}$ & $225$       & $4$ \\
        Perseus (A426)    &  15.8 &  21.2 & 14.71 & $5.6 \times 10^{-2}$ & $8.0 \times 10^{-3}$ & $3994$      & $28$ \\
        Coma (A1656)      &  14.9 &  22.0 & 14.98 & $7.9 \times 10^{-3}$ & $8.7 \times 10^{-1}$ & $256$       & $4565$ \\
        Leo (A1367)       &   4.2 &   7.4 & 14.04 & $6.2 \times 10^{-2}$ & $4.6 \times 10^{-1}$ & $3868$      & $39$ \\
        Hercules (A2199)  &  21.5 &  41.7 & 14.49 & $8.0 \times 10^{-1}$ & $8.8 \times 10^{-1}$ & $5015$      & $40$ \\
        Hercules (A2063)  &  13.5 &  33.5 & 14.46 & $6.3 \times 10^{-1}$ & $6.1 \times 10^{-1}$ & $1757$      & $6$ \\
        Abell 119         &  11.7 &  29.9 & 14.76 & $1.1 \times 10^{-1}$ & $9.7 \times 10^{-1}$ & $720$       & $540$ \\
        Abell 496         &  26.4 &  58.5 & 14.63 & $8.7 \times 10^{-1}$ & $1.4 \times 10^{-1}$ & $2642$      & $8$ \\
        Abell 1644        &  13.0 &  38.9 & 14.64 & $4.3 \times 10^{-1}$ & $8.7 \times 10^{-1}$ & $3293$      & $67$ \\
        Abell 1736        &  12.7 &  36.7 & 14.52 & $6.1 \times 10^{-1}$ & $6.6 \times 10^{-1}$ & $360$       & $51$ \\
        Shapley (A3558)   &  13.4 &  38.8 & 14.98 & $4.5 \times 10^{-2}$ & $4.0 \times 10^{-2}$ & $34\,662$   & $49$ \\
        Shapley (A3571)   &  16.5 &  54.3 & 14.61 & $8.4 \times 10^{-1}$ & $4.6 \times 10^{-1}$ & $56\,100$   & $290$ \\
        \hline
    \end{tabular}
    \caption{Properties of \texttt{SLOW} clusters from our cluster sample. We report the angular separation $\Delta\theta$ and 3D separation $\Delta r$ between \texttt{SLOW} and observed cluster positions, the \texttt{SLOW} halo mass $\log M_{500}$, the \ac{LUM} and \ac{tSZ} $p$-values, and the ratios $p^{\rm SLOW}/p^{\rm CBM}$ to the corresponding \CBM\ values from \cref{tab:cluster_properties}. Values are computed from the single \texttt{SLOW} realisation of~\citet{HernandezMartinez2024_SLOW}.}
    \label{tab:slow_plum}
\end{table*}

\end{document}